\documentclass{aa}

\usepackage{graphicx}
\usepackage{longtable}
\usepackage{soul}
\usepackage{pifont}

\usepackage{txfonts}

\usepackage[colorlinks=true, allcolors=blue]{hyperref}

\def\Mstar  {$M_{\star}$}

\def\Mdisk {$M_{\rm disk}$}

\def\Msun {$M_{\odot}$}
\def\Lsun {$L_{\odot}$}
\def\Lstar {$L_\star$}
\newcommand{\degree}{\ensuremath{^\circ}}
\def\Lline {$L_{\rm line}$}

\newcommand{\Lacc}{{$L_{\rm acc}$}}
\newcommand{\Macc}{{$\dot{M}_{\rm acc}$}}
\newcommand{\Teff}{$T_{\rm eff}$}

\newcommand{\LaccUV}{$L_{\rm acc,UV}$}
\newcommand{\LaccL}{$L_{\rm acc,line}$}
\newcommand{\Rstar}{$R_{\rm \star}$}

\usepackage{xcolor}
\DeclareUnicodeCharacter{0301}{\'{e}}
\DeclareUnicodeCharacter{0301}{\'{i}}
\DeclareUnicodeCharacter{2212}{-}

\usepackage{multirow}
\usepackage{comment}
\defcitealias{Fiorellino_2025}{F25}
\defcitealias{Alcala_2017}{A17}
\defcitealias{Baraffe_2015}{B15}
\defcitealias{Piscarreta_2025}{P25}

\usepackage{ltablex}
\keepXColumns

\begin{document}

   \title{X-shooter survey Across Regions and Ages to probe Disk Evolution (ARADE)}

   \authorrunning{Piscarreta et al.}

   \subtitle{Accretion properties in Orion A and their relation with disk masses}

   \author{L. Piscarreta\inst{\ref{ESO}}\thanks{Corresponding author \email{lara.alvopiscarreta@eso.org}}
   , 
          G. Beccari\inst{\ref{ESO}}, 
          C. F. Manara\inst{\ref{ESO}},
          R. Anania\inst{\ref{Dublin:Rossella}},
          S. E. van Terwisga\inst{\ref{Graz}},
          V. Almendros-Abad\inst{\ref{Palermo}},
          \\H. M. J. Boffin\inst{\ref{ESO}},
          R. A. B. Claes\inst{\ref{Maynooth}},
          A. Empey\inst{\ref{Dublin:Aaron}},
          B. Ercolano\inst{\ref{LMU},\ref{origins},\ref{MPE}},
          T. Jerabkova\inst{\ref{Brno}},
          K. Maucó\inst{\ref{Mexico}},
          and M. Vioque\inst{\ref{ESO}}
          }

   \institute{European Southern Observatory, Karl-Schwarzschild-Str. 2, 85748 Garching bei München, Germany\label{ESO}
         \and
         School of Physics, Trinity College Dublin, the University of Dublin, College Green, Dublin 2, Ireland\label{Dublin:Rossella}
         \and
         Space Research Institute, Austrian Academy of Sciences, Schmiedlstr. 6, 8042 Graz, Austria\label{Graz}
         \and
         Istituto Nazionale di Astrofisica (INAF)—Osservatorio Astronomico di Palermo, Piazza del Parlamento 1, 90134 Palermo, Italy\label{Palermo}
         \and
         Maynooth University Department of Physics, National University of Ireland Maynooth, Maynooth, Co. Kildare, Ireland\label{Maynooth}
         \and
         University College Dublin, Belfiled, Dublin, Ireland\label{Dublin:Aaron}
         \and
         University Observatory, Faculty of Physics, Ludwig-Maximilians-Universität München, Scheinerstr. 1, 81679 Munich, Germany\label{LMU}
         \and
         Exzellenzcluster “Origins”, Boltzmannstr. 2, 85748 Garching bei München, Germany\label{origins}
         \and
         Max-Planck-Institut für Extraterrestrische Physik, Giessenbachstr. 1, 85748 Garching bei München, Germany\label{MPE}
         \and
         Institute of Theoretical Physics and Astrophysics, Masaryk University, Kotlářská 2, Brno 611 37, Czech Republic\label{Brno}
         \and
         Universidad Nacional Autónoma de México, Instituto de Astronomía, AP 106, Ensenada 22800, BC, México\label{Mexico}
            }

   \date{Received 15 June 2026 / Accepted 27 August 2026}

  \abstract{}{We present the largest homogeneous VLT/X-Shooter study of accretion in young stellar objects spanning the entire Orion A complex. Our sample includes 91 pre-main sequence stars hosting a protoplanetary disk according to Spitzer photometry, selected to span a wide range of optical colors, of which 34 have complementary ALMA dust mass measurements.} 
   {We derived stellar and accretion properties using a self-consistent multi-component fitting procedure. Our sample encompasses spectral types from K3 down to M5, corresponding to stellar masses within $\sim$0.8$-$0.1 \Msun. }
   {The accretion-stellar luminosity (\Lacc$-$\Lstar) and the mass accretion rate-stellar mass (\Macc$-$\Mstar) scaling relations in Orion A are consistent with those of other star-forming regions (SFRs) that span a range of stellar densities, far-ultraviolet (FUV) irradiation fields, and ages, with all regions occupying the same locus in parameter space. For the 34 sources with complementary dust mass measurements, we present the first investigation of the \Macc\ and disk mass (\Mdisk) correlation in Orion A, recovering a spread consistent with that reported for other regions. Despite our sample spanning nearly five orders of magnitude in local FUV field strength, we find no statistically significant correlation between \Macc\ and FUV irradiation. However, by combining our sample with literature measurements in Lupus, the Orion Nebula Cluster, and $\sigma$ Orionis, we find a tentative population-level decrease of the inferred disk lifetime $t_{\rm disk} =$\Mdisk/\Macc\ toward stronger FUV environments, though large intrinsic scatter and limited sample sizes at intermediate and high FUV fields prevent firm conclusions.}
   {The homogeneous \Macc\ measurements reported here provide a solid foundation for future studies of accretion and disk evolution across the diverse environments of Orion A and other SFRs.}


   \maketitle

\section{Introduction}

The interaction between young stars and their circumstellar material plays a key role in shaping both stellar evolution and the early stages of planet formation. A fundamental aspect of this interaction is the accretion of gaseous material from the protoplanetary disk onto the central pre-main sequence (PMS) star (e.g., \citealt{Calvet_2000}).

In the widely accepted picture, young low-mass stars accrete through the magnetospheric accretion paradigm \citep{Koenigl_1991,Shu_1994,Hartmann_1994,Bouvier_2007}, where the disk material is funneled along the magnetic field lines of the PMS star and impacts its surface at nearly free-fall velocities producing accretion shocks (hot spots). This accretion process results in the emission of an UV/optical continuum excess as well as to enhanced emission in different lines (e.g., Balmer series; \citealt[][]{Muzerolle_2001,Calvet_Gullbring_1998,Alcala_2014}), which are otherwise mainly chromospheric in non-accreting PMS stars \citep[e.g.,][]{White_Basri_2003,Manara_2013a,Stelzer_2013}. Both of these accretion tracers are key to constrain the accretion shock luminosity, \Lacc. In turn, when combined with accurate stellar properties, this quantity allows the estimation of the mass accretion rate, \Macc\ \citep{Gullbring_1998,Hartmann_2016}. 

In previous works, the UV excess emission produced by accretion has been successfully modeled using two main approaches. The first uses hydrogen slab models in local thermodynamic equilibrium, which provide a simple, effective and widely used description of the accretion continuum (e.g., \citealt{Valenti_1993,Hartigan_2003,Manara_2013b,Claes_2024}), and the second utilizes more complex multi-component accretion-shock models (e.g., \citealt{Calvet_Gullbring_1998,Ingleby_2013,Robinson_Espaillat_2019,Espaillat_2021,Pittman_2022}), which aim to describe the physical structure of the accretion columns and shock regions. Recent studies have found that the accretion properties derived from both approaches are statistically consistent at the population level \citep[e.g.,][]{Pittman_2025,Fiorellino_2025}, suggesting that although individual measurements may vary, the simplified slab-model approach remains adequate for surveying accretion properties.
In addition to these continuum-based approaches, the luminosity of emission lines can also be used as a diagnostic of accretion, with empirical correlations between line luminosity (\Lline) and \Lacc\ enabling a straightforward conversion between the two (e.g., \citealt{Herczeg_Hillenbrand_2008,Alcala_2017,Fiorellino_2025}). Together, these methods have made possible the systematic characterization of accretion properties across large samples of PMS stars in multiple star-forming regions (SFRs), largely thanks to intermediate-resolution spectrographs such as X-Shooter at the Very Large Telescope (VLT; e.g., \citealt{Rigliaco_2012, Frasca_2017, Rugel_2018, Venuti_2019}).

The combination of these spectroscopic efforts with interferometric observations of protoplanetary disks in the radio regime with the Atacama Large Millimeter and sub-millimeter Array (ALMA) has been particularly powerful, revealing correlations between mass accretion rate, stellar mass, and disk mass. In particular, \Macc, correlates with both the stellar mass, \Mstar, and the disk mass, \Mdisk, however with large intrinsic dispersions of up to $\sim$2\,dex \citep[e.g.,][]{Hartmann_1998,Natta_2006,Manara_2016,Alcala_2014,Alcala_2017,Venuti_2019,Delfini_2025} and $\sim$1\,dex around the line \Mdisk/\Macc$\sim$1 Myr \citep[e.g.,][]{Manara_2016b,Mulders_2017,Testi_2022,Almendros-Abad_2024}, respectively. 
The origin of these spreads has been an important topic of investigation. Observational uncertainties and systematic effects such as the dependence of estimated stellar masses on the assumed evolutionary models \citep[e.g.,][]{Braun_2021,Alzate_2023,Zallio_2026}, the assumption of a single-temperature photosphere despite young stars often being heavily spotted \citep[e.g.,][]{Gully-Santiago_2017,Perez-Paolino_2023,Perez-Paolino_2024}, and the caveats associated with estimating total disk masses from dust thermal emission at a single frequency \citep[e.g.,][]{Carrasco-Gonzalez_2019,Radley_2026}, likely contribute to the measured dispersions. Nevertheless, growing evidence suggests that part of the observed spread is physical in origin reflecting differences in internal (e.g., internal photoevaporation, planet formation, \citealt{Rosotti_2017}; dust evolution, \citealt{Sellek_2020}) and external processes (e.g., dynamical interactions, \citealt[][]{Cuello_2023}; external photoevaporation, \citealt[][]{Rosotti_2017,Winter_2018}, and late-stage infall, \citealt[][]{Gupta_2024,Winter_2024}) acting differently on individual disks across a population.
The relative importance of each process, however, remains an open question. 
Accretion variability on observable timescales has been shown to contribute $\lesssim$0.5\,dex to the scatter \citep[e.g.,][]{Biazzo_2012,Costigan_2014,Venuti_2014,Claes_2022,Zsidi_2022}, and is therefore insufficient to account for the full observed dispersion.

The spectroscopic surveys of accretion previously mentioned have mainly targeted nearby ($d<$ 300\,pc), low-mass SFRs (e.g., Lupus, Taurus). Expanding the number of such surveys to denser environments and environments hosting massive stars is of particular importance to better understand the impact of the environment on disk evolution. 
In this context, the Orion complex provides the most massive and actively star-forming molecular clouds within 500\,pc of the Sun \citep{Blaauw_1964,Brown_1994,Bally_2008,Briceno_2008,Muench_2008}. In particular, Orion A ($d\sim$ 400\,pc) is the southernmost and most massive molecular cloud of the Orion complex, and can be broadly divided into a northern ``Head'' including NGC~1977, the Orion Nebula Cluster (ONC), and NGC~1980, and a southern ``Tail'', comprising the more diffuse L1641 and L1647 regions \citep[e.g.,][]{Grossschedl_2018,Kounkel_2018}. It contains a rich and diverse young stellar population (typical mean ages of $\sim$1$-$3 Myr, though local variations are present; e.g., \citealt{Hsu_2013,Getman_2014,DaRio_2016,Kounkel_2018}) that spans all evolutionary classes, from deeply embedded protostars to disk-less PMS stars, representing a valuable laboratory for the study of disk evolution \citep[e.g.,][]{Megeath_2012,DaRio_2016,Grossschedl_2019}. 

Consequently, Orion A has been the target of several spectroscopic surveys of young stellar objects (YSOs) using optical and near-infrared emission-line diagnostics (e.g., hydrogen recombination lines and Ca~{\sc ii}), which have provided valuable insights into the accretion properties of its population \citep[e.g.,][]{Hillenbrand_1997,DaRio_2010,DaRio_2016,Hillenbrand_2013,Hsu_2012,CarattioGaratti_2012,Fang_2013,Fang_2021,Kim_2016}. However, aside from a few studies of selected members in the ONC \citep{deAlbuquerque_2020}, a homogeneous survey of accretion supported by access to the Balmer jump and with a population-level coverage comparable to that achieved in nearby, less dense SFRs is still lacking.

To address this, \citet{Piscarreta_2025} (hereafter \citetalias{Piscarreta_2025}) presented a homogeneous VLT/X-Shooter analysis of 40 disk-bearing PMS stars in the Orion Nebula identified from Spitzer observations \citep{Megeath_2012}. Stellar properties, extinction values, and accretion parameters were reported for 37 of the 40 sources studied. The derived accretion parameters are similar to those reported for comparable young populations in lower-mass SFRs. In the present work, we extend the study of \citetalias{Piscarreta_2025} by presenting and analyzing 51 new VLT/X-Shooter spectra, resulting in a total sample of 91 sources. Notably, the presented sample extends beyond the Orion Nebula into the more diffuse ``Tail'' of the Orion A complex (L1641/L1647). Paired with available ALMA millimeter observations, this allows us, for the first time, to examine how mass accretion rates relate to the available disk mass across this environment.

This paper is the first in a series called ARADE (x-shooter survey Across Regions and Ages to probe Disk Evolution), which aims to characterize the stellar and accretion properties of young stars across a range of environments and ages in a homogeneous way. Building on existing VLT/X-Shooter surveys of nearby regions such as Lupus, Chamaeleon I, and Upper Scorpius, ARADE extends this self-consistent baseline to two complementary populations analyzed with the same methods: Orion A, a young ($\sim$1–3 Myr) and comparatively massive, dense environment, and the Upper-Centaurus-Lupus and Lower-Centaurus-Crux subgroups of Scorpius–Centaurus, which trace a more evolved stage of disk evolution. In this first paper, we focus on the accretion properties of young stars located across the entire Orion A complex.

\section{Dataset}\label{sec:data}

\subsection{Sample selection}\label{sec:sample}

\begin{figure}[t]
\centering
    \includegraphics[width=0.45\textwidth]{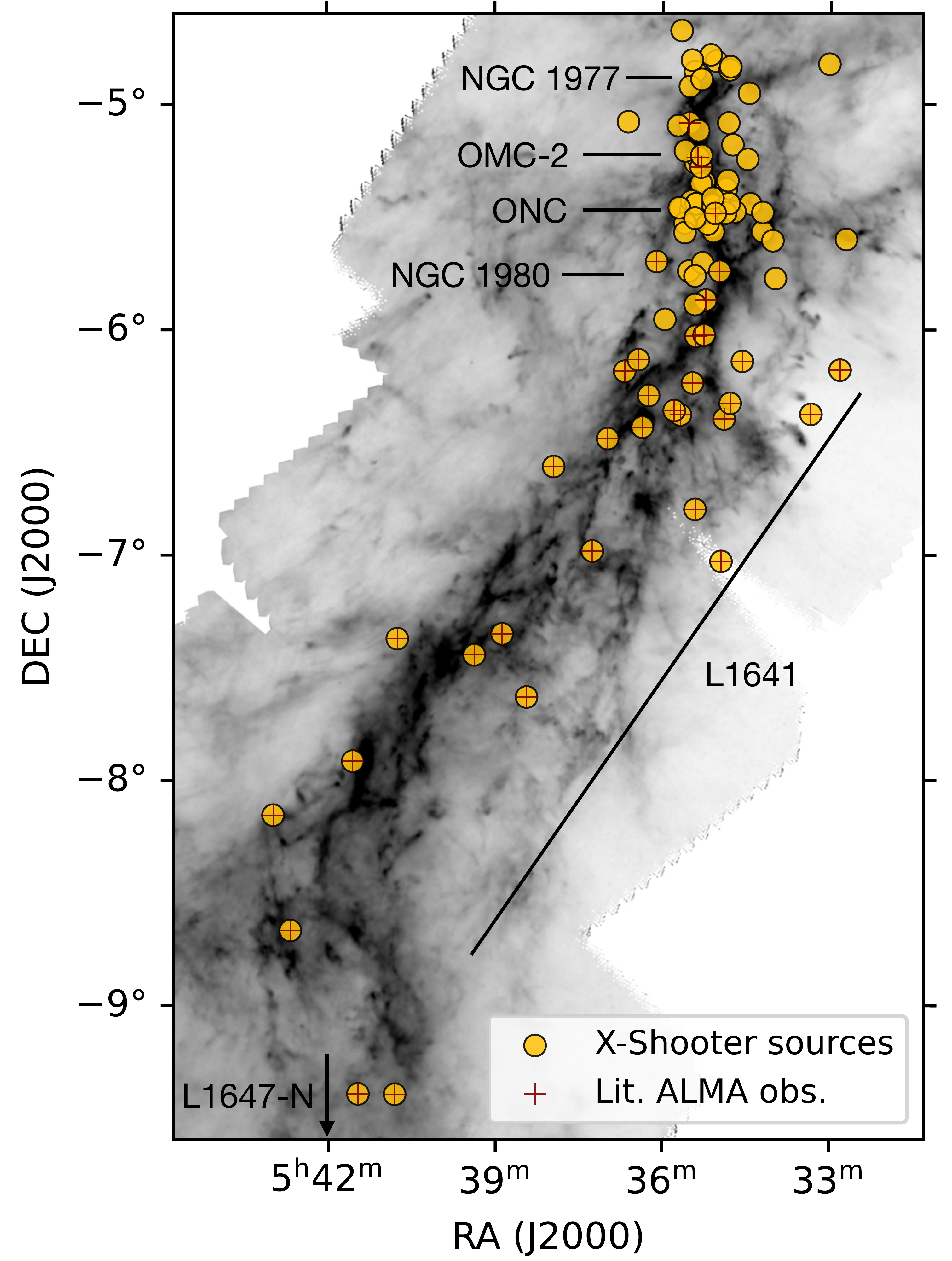}
\caption{Spatial distribution of our targets across the Orion A molecular cloud. Sources with dust disk mass measurements available from the literature (App.~\ref{app:dust_masses}) are marked with red plus symbols. The grayscale background shows the gas column density map from the Herschel Gould Belt Survey.}
\label{fig:loc}     
\end{figure}

In this work, we extend the analysis presented by \citetalias{Piscarreta_2025} by increasing the number of PMS stars analyzed and extending the coverage to the entire Orion A complex, from NGC 1977 down to the L1641 and L1647 clouds (Fig.~\ref{fig:loc}). The new 51 PMS stars presented in this work were selected considering the following criteria: 

\begin{itemize}
    \item Objects classified by \citet{Megeath_2012} as ``Disk'' according to Spitzer photometry. In their catalog, YSOs with an IR excess not classified as a protostar were assigned to the ``disk'' class, meaning that transition disks and optically thick inner disks were not distinguished;
    \item Relative Gaia DR3 parallax error less than 10\%. We note that, although this criterion was applied to the newly presented sources, it was not imposed in the dataset presented by \citetalias{Piscarreta_2025} (further described below);
    \item Sources brighter than $r =$17.5\,mag;
\end{itemize}

From the stars satisfying the above criteria, we selected PMS stars spanning the full range of $r-i$ colors in the OmegaCAM color-magnitude diagram (CMD) within $r=$13.5$-$17.5\,mag, corresponding to ages from $<$1 to $>$10 Myr based on the PISA stellar evolutionary models (\citealt{PISA_isochrones}; Fig.~2 of \citetalias{Piscarreta_2025}). The isochrones were shifted onto the observed CMD adopting a fixed distance of 395\,pc, $E(B-V)$ = 0.052\,mag \citep{Jerabkova_2019}, and $R_V$ = 3.1\,mag. The adopted distance agrees well with the sample's Gaia DR3 parallax distribution, $\varpi = 2.51\pm$0.09\,mas ($d\sim$398$\pm$14\,pc), and the commonly adopted distance to Orion A \citep[e.g.,][]{Kounkel_2018}.

The sample selection prioritized disk presence, and broad spread in accretion signatures (through the H$\alpha-$r vs. $r-i$ color-color diagram) and $r-i$ color, rather than Gaia astrometric quality indicators such as the Re-normalized Unit Weight Error (RUWE) or the fidelity parameter, {\tt fidelity\_v2} \citep{Rybizki_2022}. Well-behaved single-star solutions are typically taken as RUWE$\lesssim$1.4 \citep{Lindegren_2021}, while higher values might suggest binarity or other perturbing effects \citep{Lindegren_2021}. Similarly, {\tt fidelity\_v2} $>$ 0.5 is generally adopted to indicate a reliable astrometric solution.
Approximately 20\% of sources have RUWE $>$ 1.4, which is not unexpected given that the presence of a protoplanetary disk is known to inflate the RUWE parameter \citep{Fitton_2022}. Adopting the RUWE $>$ 2.5 threshold more appropriate for disk-bearing young stars reduces this fraction to $\sim$8\%. Similarly, only $\sim$7\% of sources have {\tt fidelity\_v2} $<$ 0.5, and only three sources are flagged by both criteria. These are part of the sample presented by \citetalias{Piscarreta_2025} and show relative parallax errors exceeding 10\%. We flag these sources in Table~\ref{tab_app:gaia_info} and note that lithium is detected in the three spectra, confirming their young nature. 

For the YSOs in the L1641 and L1647 clouds, we made sure that the targets had a counterpart observation from the Survey of Orion Disks with ALMA \citep[SODA;][]{vanTerwisga_2022}, in order to enable a joint analysis of accretion and disk properties. SODA surveyed the unresolved millimeter continuum emission of $>$800 protoplanetary disks in the southern part of Orion A with ALMA, providing dust mass estimates for $\sim$500 of them.
Additionally, a few other sources have complementary ALMA continuum observations available in the literature. 
In total, dust mass measurements from the literature are available for about one third of our sample composed of 91 PMS stars (App. \ref{app:dust_masses}). Figure \ref{fig:loc} summarizes the spatial distribution of the Orion A PMS stars analyzed atop the gas column density map from the Herschel Gould Belt survey\footnote{\url{http://www.herschel.fr/cea/gouldbelt/en/Phocea/Vie_des_labos/Ast/ast_visu.php?id_ast=66}} \citep[][]{Andre_2010,Roy_2013,Polychroni_2013}. The figure already hints at the scarce coverage of millimeter observations in the Orion Nebula region, a limitation that will be further discussed in Sect.~\ref{subsubsec:environment}.

\subsection{Observations and data reduction}

We observed the 91 targets using the intermediate-resolution X-Shooter spectrograph located in the Unit Telescope (UT) 3 of the Very Large Telescope (VLT) in Paranal, Chile \citep{Vernet_2011}. With its three arm (UVB, VIS, NIR), the VLT/X-Shooter covers a wide effective wavelength range from 300 to 2500\,nm and is, therefore, uniquely suited to characterize accreting young stars.
The spectra analyzed were collected along three programs: the first conducted between October 2021 and January 2022 (program ID 108.2206), which targeted YSOs in the Orion Nebula (data analyzed by \citetalias{Piscarreta_2025}); the second between October 2024 and March 2025 (program ID 114.276M); and the third between February and March 2026 (program ID 116.2934). The last two programs expanded the statistics in the Orion Nebula and extended the coverage from the first one to the southern regions of the Orion A complex, namely the L1641 and L1647 clouds.

For the data acquired in the second and third observing programs, we adopted the same observing strategy and reduction process as that presented by \citetalias{Piscarreta_2025}, where full details are given. In brief, the observations were performed in slit-nodding mode considering an ABBA cycle and slit widths of 1.3'', 0.9'', and 0.9'' for the UVB, VIS, and NIR arms, respectively, resulting in spectral resolutions of R$\sim$4100, 8900, and 5600. A short exposure using a 5.0''-slit in stare mode was also included immediately after each science exposure in all three X-Shooter arms, to account for potential slit losses on the narrower-slit observations and allow for flux calibration as described by \citet{Manara_2021}.

The observing conditions were generally good, with most of the sources being observed in clear or photometric conditions. 
The median seeing during the observations was 0\farcs9, with an inter-quartile range of 0\farcs8$-$1\farcs1. We note that source \#57 was observed under poorer seeing conditions (seeing $\sim$2\farcs2), although sky conditions were clear. We verified its flux calibration by comparing synthetic photometry derived from the X-Shooter spectrum with archival photometry from Gaia DR3 and \textit{VISTA}. These agree within $\lesssim$0.4\,mag in all bands except $H$, where the difference approaches 1\,mag. Given the non-simultaneity of the observations, and the absence of similarly large discrepancies in adjacent bands, we do not consider this difference indicative of a significant flux-calibration issue. In any case, the stellar and accretion properties derived in this work rely on the UVB/VIS portion of the X-Shooter spectrum, where the synthetic photometry and archival photometry are in good agreement.

We report the observations log in Table~\ref{tab:log_obs} along with the estimated S/N at the wavelength ranges [399, 402]\,nm, [698, 702]\,nm, and [1648, 1652]\,nm for the UVB, VIS, and NIR arms, respectively.
The data reduction and telluric correction were performed using the X-Shooter pipeline \citep{xshooter_pipeline} and {\tt Molecfit} \citep{Smette_2015,Kausch_2015}, respectively, both within the {\tt EsoReflex} environment \citep{esoreflex}.

\section{Analysis and results}\label{sec:results}

To accurately characterize the stellar properties as well as the accretion parameters in YSOs, it is important to simultaneously model the photospheric emission, the  accretion continuum excess, and extinction, the latter of which may arise from both the interstellar medium and circumstellar material. This simultaneous multi-component fitting procedure is crucial since accretion and extinction have degenerate effects on the spectra of YSOs. 

We used \texttt{FRAPPE}\footnote{\url{https://github.com/RikClaes/FRAPPE}} (FitteR for Accretion ProPErties; \citealt{Claes_2024}), which is based on the original method presented by \citet{Manara_2013b}. \texttt{FRAPPE} considers a set of interpolated Class III YSO templates ranging from G8 down to M9.5\footnote{We note that the interpolated grid adopted in \texttt{FRAPPE} considers half-integer SpTs that might not be fully represented in observed standard star spectra \citep{MaizApellaniz_2025}.} to account for the photospheric emission from the central star, a grid of hydrogen slab models to model the accretion contribution while the extinction ($A_V$) is taken as a free parameter. Here we present the stellar and accretion parameters estimated using this method. We also present accretion luminosities derived by implementing empirical relations with respect to line luminosities.

\subsection{Stellar properties and accretion parameters from UV excess}\label{subsec:frappe_outputs}

The core idea behind \texttt{FRAPPE} is to determine the optimal combination of spectral type (SpT), $A_V$, and the parameters describing the accretion continuum that minimizes a $\chi^{2}_{\,\text{like}}$ metric \citep{Manara_2013b,Claes_2024}.
We considered a grid of extinction values ranging from 0 to 5\,mag with a step of 0.1\,mag, applying the extinction law of \citet{Cardelli_1989} with $R_{V}=$3.1\,mag to de-redden the observed spectra. The typical uncertainty assumed on $A_V$\ is 0.2\,mag \citep{Claes_2024}.
The best fit SpT is assumed to be that of the best fit interpolated Class III template with typical uncertainties of $\pm$0.5 subtypes for M-types and $\pm$1.0 subtype for earlier SpTs \citep{Manara_2017a}. We compared the SpTs that we derived with SpT estimations available in the the literature (Table~\ref{tab:fitter_output} compiles literature SpTs and the new measurements presented in this work). Our estimates are in overall good agreement with the ones previously published for the majority of our sources, with $\sim$76\% of the sources agreeing within $\pm$1 SpT and $\sim$92\% within $\pm$2 SpTs.

The effective temperature ($T_{\rm eff}$) is retrieved by implementing the SpT$-$\Teff\ relations from \citet{Herczeg_Hillenbrand_2014}. Their bolometric correction is also used (considering the revision presented by \citealt{Claes_2024} for stars with \Teff$<$4500 K) to obtain the bolometric flux, which in turn is converted into stellar luminosity (\Lstar) from \Lstar$=$4$\pi$d$^{2}F_{bol}$. Distances to individual targets were taken from \citet{Bailer-Jones_2021}, which are consistent with simple parallax inversion for all sources in our sample (the latter was used by \citetalias{Piscarreta_2025}). Using \citet{Bailer-Jones_2021} distances here results in only minor changes to the parameters for targets in common between the two samples, well within their respective uncertainties.
Uncertainties on \Teff\ were obtained by propagating the SpT uncertainty through the SpT$-$\Teff\ relation of \citet{Herczeg_Hillenbrand_2014}, while the uncertainty on \Lstar\ is assumed to be 0.2\,dex in line with previous VLT/X-Shooter studies (e.g., \citealt{Rigliaco_2012,Alcala_2014,Manara_2017a}).

While \texttt{FRAPPE} supports multiple evolutionary models (e.g., \citealt{Baraffe_2015} and \citealt{Feiden_2016}), we adopted the PISA isochrones \citep{PISA_isochrones} for consistency with \citetalias{Piscarreta_2025} and because they were specifically computed for the stellar population in Orion \citep{Jerabkova_2019}. In practice, PISA provides mass and age estimates for all but one source in our sample (source \#57), whereas, for example, \citet{Baraffe_2015} leaves a larger number of sources outside the model grid. The overall properties of the population would not differ substantially by adopting \citet{Baraffe_2015} instead. We note that the choice of evolutionary tracks remains an open debate in the field \citep[e.g.,][]{Zallio_2026,Towner_2026}, though a detailed assessment is beyond the scope of this work.

Stellar masses (\Mstar) and ages were then obtained by interpolating each target's location on the Hertzsprung-Russell diagram (HRD).
More specifically, we first interpolated the set of PISA isochrones onto a finer 1000$\times$1000 grid in age and mass. Then, for each target, we performed a Monte Carlo simulation with 1000 iterations, randomly sampling \Teff\ and \Lstar\ from Gaussian distributions consistent with their uncertainties, and selecting the age and mass closest grid point in the interpolated plane. The adopted mass and age are the medians of the resulting distributions, with uncertainties from the 25th and 75th percentiles.

\begin{figure*}[h!]
\centering
    \includegraphics[width=0.47\textwidth]{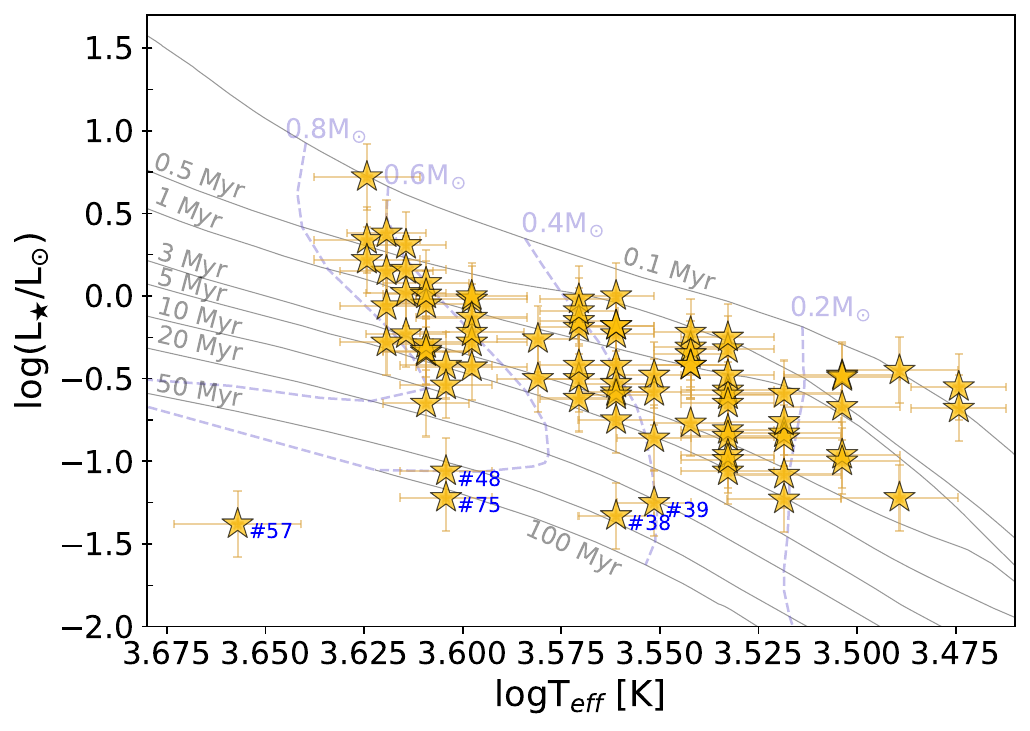}
    \includegraphics[width=0.47\textwidth]{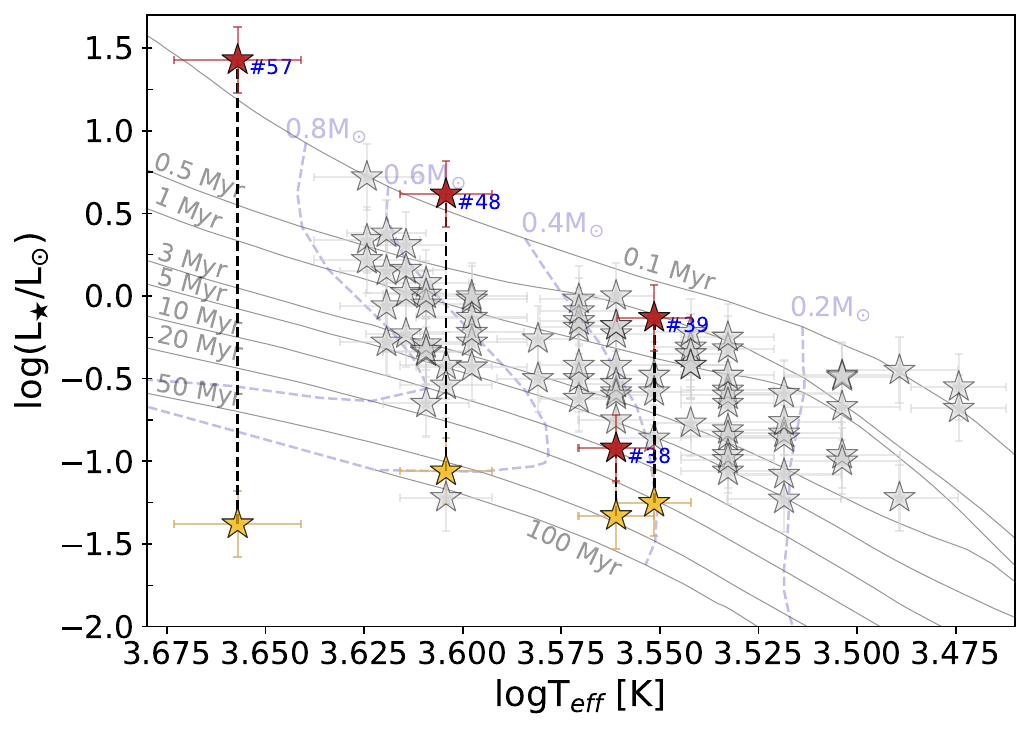}
\caption{\textit{Left panel:} HRD diagram constructed with the \Teff\ and \Lstar obtained with \texttt{FRAPPE}. The black solid lines show the PISA isochrones \citep{PISA_isochrones}, as labeled. The mass tracks are presented as dashed purple lines. Sub-luminous sources are identified by their IDs. \textit{Right panel:} HRD focused on the sub-luminous sources, showing their positions before (yellow stars) and after (brown stars) correction for the assumptions described in Sect.~\ref{subsec:sub-luminous}.}
\label{fig:HRD}      
\end{figure*}

We were able to successfully run \texttt{FRAPPE} for 84 spectra out of the total 91 spectra included in our sample. The seven sources for which we do not report outputs from the multi-component fitting procedure are \#9, \#28, \#32, \#42, \#44, \#54, and \#60 on Table~\ref{tab_app:gaia_info}. Source \#60 (V2331 Ori) is in our sample because it was classified as a PMS star with a disk by \citet{Megeath_2012}, but subsequent studies at longer wavelengths report this source as a protostar with a Flat SED evolutionary class (e.g., \citealt{Tobin_2020,Federman_2023}). For all the other cases, the photospheric continuum of the observed spectra cannot be accurately traced (due to very low S/N, strong veiling or emission lines), preventing a reliable solution with \texttt{FRAPPE}.

Figure \ref{fig:HRD} shows the HRD for our sample, which comprises sources with estimated \Teff$\sim$4500$-$3000 K (corresponding to SpTs$=$K3$-$M5) and \Mstar$\sim$0.1$-$0.8 \Msun. The estimated stellar luminosities range from 0.04\Lsun\ and $\sim$5L$_{\rm \odot}$ with source \#57 as the less luminous and source \#66 as the most luminous. Most sources are located in the region of the HRD consistent with isochronal ages between 0.5 and 5 Myr. However, five targets show significantly lower \Lstar\ for a fixed \Teff\ with respect to the bulk population ($\Delta$log\Lstar$\sim$1\,dex). These sources are \#38, \#39, \#48, \#57, and \#75, and will be discussed further in Sect.~\ref{subsec:sub-luminous}.
Even when not accounting for these sub-luminous sources, there is still a considerable spread in \Lstar\ at fixed \Teff. Such substantial luminosity dispersion have been observed in nearly all nearby SFRs \citep[e.g.,][]{Hillenbrand_2008,Reggiani_2011} and it is thought to arise from a combination of starspots, accretion history, binarity, and disk orientation \citep{Gully-Santiago_2017,Baraffe_2017,Guarcello_2010}, rather than measurement or age uncertainties alone \citep[e.g.,][]{Fang_2021}.
The youth of our sample (and PMS nature) is further corroborated by the presence of the lithium absorption feature at $\lambda$670.8\,nm, detected in all sources for which \texttt{FRAPPE} could be applied (including the sub-luminous sources). The only sources with a non-detection are \#9, \#32, and \#44, all sources for which we do not report a \texttt{FRAPPE} fit. While the first two were already discussed by \citetalias{Piscarreta_2025}, the non-detection in \#44 is likely due to high veiling filling in the absorption feature rather than a genuine absence of lithium, as the spectrum is dominated by strong emission lines.

As anticipated, alongside extinction and the stellar properties, \texttt{\texttt{FRAPPE}} derives simultaneously an estimate of the accretion luminosity.
The accretion luminosity, \Lacc, is obtained by integrating the best-fitted slab model over its entire wavelength range (50$-$2500\,nm) with typically assumed uncertainties of 0.25\,dex. The mass accretion rate, \Macc, is subsequently derived using \Macc $= 1.25 \times$ \Lacc \Rstar/(G \Mstar) from \citet{Hartmann_2016}, adopting the stellar radius and mass inferred respectively from the Stefan-Boltzmann equation and the location of each source on the HRD relative to evolutionary models. 
We list in Table~\ref{tab:fitter_output} the stellar and accretion properties obtained with \texttt{FRAPPE}. As already mentioned, we adopted the PISA estimates throughout this work, but also report the \citet{Baraffe_2015} values for completeness and ease of comparison with studies using different isochrone sets.

\subsection{Line-based accretion luminosities}\label{subsec:Lacc_lines}

Modeling the UV excess emission provides the most direct estimate of \Lacc\ generated from the accretion shock regions. However, many permitted emission lines spanning from UV to IR wavelengths arise in the magnetospheric accretion columns and are hence signatures of ongoing mass accretion \citep{Calvet_Gullbring_1998,Muzerolle_1998a,Muzerolle_1998b}, with H$\alpha$ emission in particular having been used to directly model the magnetospheric accretion flow \citep{Pittman_2025}. As a result, the luminosities of these lines are expected to correlate with \Lacc\, as it has been robustly demonstrated observationally \citep[e.g.,][]{Natta_2004,Herczeg_Hillenbrand_2008,Rigliaco_2012,Alcala_2014,Alcala_2017,Fiorellino_2025}. We derived \Lacc\ from different accretion line diagnostics and compare these estimates with those obtained from the UV excess fitting procedure. 

Each spectrum was de-reddened assuming the extinction derived with \texttt{FRAPPE}. The local continuum around each diagnostic line is modeled using a third-degree polynomial, iteratively fitted with a 2$\sigma$-clipping procedure to exclude the line itself and any other neighboring features. After subtracting the continuum, the line flux, $F_{line}$, is measured by direct integration of the residual line profile. Line luminosities are then obtained using $L_{line} = 4\pi d^{2} F_{line}$, assuming distances from \citet{Bailer-Jones_2021} and uncertainties derived through standard error propagation. We converted the measured line luminosities into accretion luminosities using the empirical \Lacc $-$ \Lline\ relations from \citet{Alcala_2017} (hereafter \citetalias{Alcala_2017}) and their recent update by \citet{Fiorellino_2025} (hereafter \citetalias{Fiorellino_2025}). 

We adopted the \Lacc $-$ \Lline\ calibrations for H$\alpha$, H$\beta$, H$\gamma$, H$\delta$, Pa$\beta$, Pa$\gamma$, Br$\gamma$, \ion{He}{i} $\lambda$587\,nm, \ion{He}{i} $\lambda$667\,nm, and Ca~{\sc ii} K emission lines. The final \LaccL\ value for each target was computed as the mean across all available lines and the respective error as the standard deviation. We compared the \LaccL\ values obtained with both set of empirical relations to the \LaccUV\ value derived with \texttt{FRAPPE} and found that the slopes are consistent with unity for both cases, indicating an overall agreement between both accretion luminosity estimates. However, while the \citetalias{Alcala_2017} calibration shows an intercept consistent with zero, the \citetalias{Fiorellino_2025} relations display a positive offset of $\sim$0.22, indicating systematically higher accretion luminosity values compared to the UV-based estimates. 
As also pointed out by the authors, this systematic difference likely reflects the distinct wavelength coverage, the model used to obtain the \Lacc\ reference from UV fitting, and the heterogeneity of SFRs considered. Specifically, the \citetalias{Fiorellino_2025} relations were re-calibrated using data from the ULLYSES (UV Legacy Library of Young Stars as Essential Standards) program \citep{Roman-Duval_2020,Roman-Duval_2025} extending therefore the UV coverage below $\sim$300\,nm. Moreover, the calibration of these updated relations adopted the accretion shock model to estimate \LaccUV\ \citep{Calvet_Gullbring_1998,Pittman_2022,Pittman_2025}, and were based on the PENELLOPE sample \citep{Manara_2021}, which spans a broader range of SFRs (e.g., Orion OB1, $\sigma$ Ori) compared to the Lupus-focused sample of \citetalias{Alcala_2017}. These aspects represent methodological improvements providing a more comprehensive and broadly applicable calibration, and we therefore adopted the \citetalias{Fiorellino_2025} estimates throughout the remainder of the analysis.

\subsection{Sub-luminous sources}\label{subsec:sub-luminous}

Once we accounted for extinction and accretion, and determined accurate stellar properties for the sample, we found five sub-luminous sources in the HRD (Fig.~\ref{fig:HRD}; sources \#38, \#39, \#48, \#57, and \#75). 
Their young nature is nevertheless confirmed by the detection of lithium in all six sources, suggesting they are genuinely young objects. A plausible explanation is that their disks are viewed at close to edge-on inclinations, which can suppress the stellar flux and mimic sub-luminous photospheres in the HRD.
This concept is meaningful under the assumption of gray obscuration (i.e., wavelength-independent obscuration), in which the stellar and accretion continuum are attenuated without equally affecting emission lines arising from more extended regions in the disk. Such gray-like extinction is typically associated with obscured disks such as dippers \citep[e.g.,][]{Empey_2025} or major dimming events (e.g., caused by a warped inner disk; \citealt{Bouvier_2007_AATau,Facchini_2016,Schneider_2018}).

A useful diagnostic to test whether disk obscuration could potentially explain the observed $\gtrsim$1\,dex sub-luminosity relative to the bulk population is the [\ion{O}{i}]\,$\lambda$6300 $\AA$ emission line \citep[e.g.,][]{Alcala_2014,Alcala_2020}. While the UV continuum excess emission in which we base our \Lacc\ estimates is emitted in the accretion shock regions at the stellar surface \citep{Calvet_Gullbring_1998}, the [\ion{O}{i}] line is thought to originate in a more spatially extended region above the disk surface \citep[e.g.,][]{Simon_2016,Banzatti_2019,Birney_2024}. This is particularly true for the low-velocity component (LVC) of the line as any high-velocity component (HVC), if present, is typically associated with collimated jets and may therefore be subject to stronger inner disk obscuration in inclined disks \citep{Natta_2014,Nisini_2018}, similarly to the UV excess coming from accretion. By comparing the accretion luminosity derived from the LVC of the [\ion{O}{i}] line luminosity with that obtained from UV modeling, we can test whether the two estimates are in disagreement, which could suggest the presence of an inclined disk.

One possible caveat in this method is the fact that if a star is strongly affected by external photoevaporation, the [\ion{O}{i}] LVC line luminosity is no longer expected to correlate with \Lacc, as it would instead be driven by the UV flux from nearby massive stars \citep[e.g.,][]{Ballabio_2023,Mauco_2025}. We anticipate here that none of the sub-luminous sources reside in regions affected by strong external FUV fields (Sect.~\ref{subsubsec:environment}).

We implemented the multi-component Gaussian fitting from \texttt{STAR-MELT} \citep{Campbell-White_2021} to the [\ion{O}{i}] line profiles of the six sub-luminous sources. The [\ion{O}{i}] line profile of source \#39 is better reproduced by two Gaussians, whereas for sources \#48 and \#57 one single component is enough, and source \#38 needs three Gaussians. We show the fits retrieved with \texttt{STAR-MELT} in Fig.~\ref{fig:star-melt_subl}. The line is contaminated with sky in source \#75. We therefore excluded the latter source from this particular exercise on disk geometry. 

\renewcommand{\arraystretch}{1.4}
\begin{table}[t]
\caption{Accretion and stellar properties estimates retrieved under the high-inclination scenario.}
\vspace{-20pt}
\begin{center}
\scalebox{0.8}{
\begin{tabular}{lcccc}
\hline \hline
ID & \#38 & \#39 & \#48 & \#57 \\
\hline
log($L_{\rm acc}/L_\odot$) UV      & $-$1.83 & $-$1.26 & $-$1.69 & $-$3.18 \\
log($L_{\rm acc}/L_\odot$) [\ion{O}{i}]    & $-$1.43 & $-$0.14 & $-$0.01 & $-$0.38 \\
Obs. factor [dex]                  &    0.41 &    1.12 &    1.68 &    2.81 \\
log($L_{\star}/L_\odot$) UV        & $-$1.33 & $-$1.25 & $-$1.06 & $-$1.38 \\
log($L_{\star}/L_\odot$) corr.     & $-$0.92 & $-$0.13 &    0.62 &    1.43 \\
\hline
Mass corr. [$M_\odot$]                   & $0.45^{+0.08}_{-0.07}$ & $0.32^{+0.04}_{-0.04}$ & $0.52^{+0.07}_{-0.05}$ & $1.43^{+0.38}_{-0.30}$ \\
log$\dot{M}_{\rm acc}$ corr. [M$_\odot\,{\rm yr}^{-1}$] & $-8.95^{+0.27}_{-0.30}$ & $-7.81^{+0.27}_{-0.29}$ & $-8.19^{+0.27}_{-0.29}$ & $-9.82^{+0.27}_{-0.32}$ \\
\hline
\end{tabular}}
\end{center}
\vspace{-10pt}
\label{tab:subl_obs_factor}
\end{table}

After correcting the [\ion{O}{i}] line fluxes for extinction for the four sub-luminous sources inspected, we converted the line fluxes into luminosities, restricting the flux computation to the LVC (|v|$<$40\,km\,s$^{-1}$, following \citealt{Nisini_2018}).
We then implemented the \Lacc$-L_{\rm [O~I],LVC}$ relation also from \citet{Nisini_2018}. Table \ref{tab:subl_obs_factor} summarizes the different accretion luminosities obtained with \texttt{FRAPPE} and the [\ion{O}{i}].
The \Lacc\ values derived from the [\ion{O}{i}] line luminosity are significantly larger ($>$1\,dex for all sources except source \#38 with $\sim$0.4\,dex) than the values estimated from the UV excess.
As previously explained, under the assumption of a gray extinction, we corrected the stellar luminosity of the sub-luminous sources by the extra accretion luminosity detected from the [\ion{O}{i}] line flux (``obscuration factor'' on Table~\ref{tab:subl_obs_factor}).

The right panel of Fig.~\ref{fig:HRD} shows the corrected positions of the sub-luminous sources in the HR diagram (brown stars), obtained after applying obscuration factors to the stellar luminosities. The effective temperature is assumed to remain unaffected, since gray extinction uniformly suppresses the stellar flux without altering the spectral features used to derive \Teff. 
The location on the HRD of the sub-luminous sources after the correction is more in agreement with the locus occupied by young PMS stars in Orion A. Interestingly, source \#57 exhibits the largest obscuration factor ($\gtrsim$2\,dex), exceeding even Par-Lup3-4 in Lupus, a well-studied nearly edge-on disk system (for which stellar properties can still be retrieved) that appears sub-luminous by $\sim$4 magnitudes ($\sim$1.6\,dex; inclination $\sim$81$\degree$; \citealt{Comeron_2003,Huelamo_2010,Alcala_2014}). This study consistently shows that \Lacc\ derived from [\ion{O}{i}] exceeds that derived from the UV excess for all four sub-luminous sources. The correction factors themselves should be regarded as order-of-magnitude estimates rather than precise determinations, as robust disk geometry constraints would ultimately require spatially resolved observations, which are not available for these sources.

Nevertheless, this interpretation is supported by independent diagnostics. \citet{Erkal_2022} predicted that the \ion{He}{i}\,$\lambda$10830\,\r{A} line should appear purely in emission at high inclinations, as the disk obscures the stellar photosphere and suppresses the usual red- and blue-shifted absorption components, leaving only jet/wind emission. Consistently, all sub-luminous sources except \#75 show this line in emission without absorption features. The same holds for seven additional non-subluminous objects, two of which (\#9 and \#44) lack converged \texttt{FRAPPE} fits due to strong veiling, which may also mask absorption signatures. Furthermore, source \#38 was previously identified as sub-luminous by \citet{Fang_2017}, who inferred a disk inclination of $\sim$79$-$82$\degree$ from SED modeling. 
With this in mind, we adopted the revised HRD positions for sources \#38, \#39, \#48, and \#57 for the remainder of the analysis, along with the correspondingly updated stellar and accretion properties. Source \#75 is instead kept at its original position. The stellar and accretion properties under the high-inclination scenario are reported in Table~\ref{tab:subl_obs_factor}, while their original estimates from \texttt{FRAPPE} (Sect.~\ref{subsec:frappe_outputs}) are reported in Table~\ref{tab:fitter_output} together with the rest of the sample.

\subsection{Accretion luminosity detection limits}\label{subsec:acc_limits}

Young low-mass stars have active chromospheres whose emission overlap with typically used accretion diagnostics (e.g., UV excess, Balmer lines, Ca II IR triplet; \citealt{Ingleby_2011,Manara_2013a,Manara_2017}). While this contribution is negligible in strong accretors, it becomes increasingly important in the low accretion regime. Therefore, assessing the reliability of each diagnostic is essential when interpreting accretion luminosities. 

Two thresholds can be used to determine whether accretion is measurable or detected. The first concerns the contrast to which the UV-excess can be accurately measured \citep{Claes_2024} and the second represents the physical lower limit to line-based accretion estimates, below which the emission is dominated by chromospheric activity \citep{Manara_2017}. Both limits are typically referred to as accretion noise, $L_{\text{acc, noise}}$.

\begin{figure}[t]
\centering
    \includegraphics[width=0.45\textwidth]{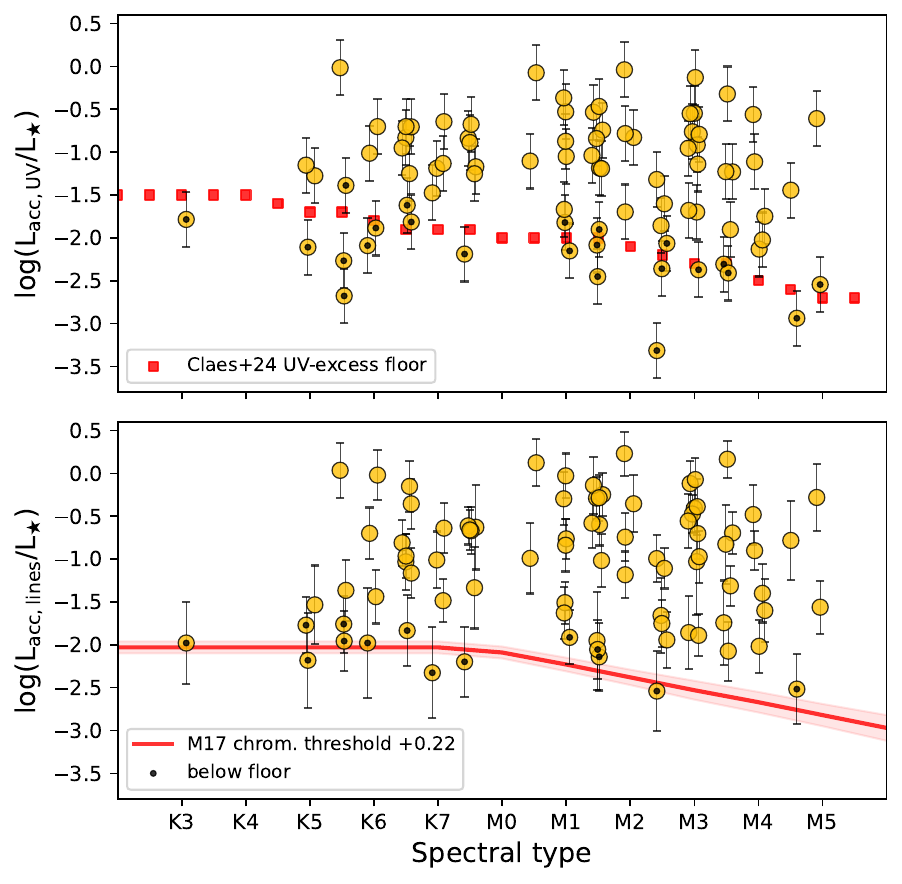}
\caption{log(\Lacc/\Lstar) as a function of SpT, showing the accretion noise limits used to identify upper limits. \textit{Upper panel:} UV-excess floor from \citet{Claes_2024}. \textit{Lower panel:} Chromospheric noise threshold from \citet{Manara_2017}, shifted by $+$0.22~dex to the \citetalias{Fiorellino_2025} scale (Sect.~\ref{subsec:Lacc_lines}). Sources consistent with lying below a threshold within their uncertainties are marked with a black dot. Twelve sources fall below both limits and are classified as accretion upper limits. Sources below only one limit are further assessed using the \ion{He}{i}\,$\lambda$10830\,\r{A} line profile and H$\alpha$ width (Sect.~\ref{subsec:acc_limits}). A small horizontal offset has been applied to avoid crowding.}
\label{fig:low_accretors}       
\end{figure}

Figure \ref{fig:low_accretors} flags accretion upper limits using both criteria. The upper panel shows log(\LaccUV/\Lstar) as a function of SpT from \texttt{FRAPPE}, with the \citet{Claes_2024} UV-excess floor shown as a dotted red line. The bottom panel of the same figure presents the \LaccL\ inferred from emission lines normalized by the \Lstar\ obtained with \texttt{FRAPPE}.
We adopted the chromospheric accretion boundary of \citet{Manara_2017}, originally defined using the \citet{Alcala_2014} \Lacc$-$\Lline\ relations, shifted by $+$0.22\,dex to place it on the \citetalias{Fiorellino_2025} scale. Since the \citetalias{Alcala_2017} and \citet{Alcala_2014} calibrations are mutually consistent, this offset is equivalent to the systematic difference between the \citetalias{Fiorellino_2025} and \citetalias{Alcala_2017} relations.

Sources lying above both boundaries by more than 1$\sigma$ are considered bona-fide accretors. Conversely, sources lying below both boundaries by more than 1$\sigma$ are classified as accretion upper limits. These are 12 out of the full sample which are, in turn, also considered upper limits in mass accretion rate for the remainder of the analysis. 
Additionally, there are 13 sources lying above only one of the two accretion noise boundaries (sources \#11, \#19, \#29, \#35, \#40, \#45, \#71, \#75, \#78, \#80, \#81, \#87, and \#90). For these, we performed a multi-tracer classification to assess their accretion status. We first checked for agreement between the presence of redshifted absorption in the \ion{He}{i}\,$\lambda$10830\,\r{A} line profile \citep{Thanathibodee_2022, Erkal_2022} and the H$\alpha$ full width at 10\% of the peak intensity, using the 270 km~s$^{-1}$ threshold of \citet{White_Basri_2003}. When both tracers agreed, we adopted that classification directly. When they disagreed or either were ambiguous, we applied a hierarchical approach: we first relied on the \ion{He}{i}\,$\lambda$10830\,\r{A} profile, as redshifted absorption features are the most unambiguous signature of magnetospheric infall. In cases where the \ion{He}{i}\,$\lambda$10830\,\r{A} profile remained inconclusive, we fell back on the UV continuum excess limit \citep{Claes_2024}. Following this procedure, we classify sources \#40, \#71, \#75, \#78, \#80, and \#81 as accretors, and sources \#11, \#19, \#29, \#35, \#45, \#87, and \#90 as upper limits. Combined with the 12 sources lying below both accretion noise limits, this results in a total of 19 accretion upper limits in our sample.
We caution here that these accretion detection limits are based on a single X-Shooter snapshot of each source's accretion activity. Therefore, some sources classified here as accretion upper limits (particularly those close to the chromospheric thresholds) could be classified as accretors at a different epoch, and vice-versa \citep[e.g.,][]{Bayo_2012}.

\section{Discussion}\label{sec:discussion}

Here we discuss the stellar, accretion, and disk properties for the sources in our sample. We compare them with those obtained in several other nearby SFRs compiled by \citet{Manara_2023}, namely, Lupus, Taurus, Chamaeleon I/II, $\rho$ Ophiuchus, Corona Australis, and Upper Scorpius, as well as those from \citet{Mauco_2023} for $\sigma$~Orionis ($\sigma$~Ori). Orion~A offers a unique perspective in this context, as it encompasses a wide diversity of environments within a single star-forming complex at young ages ($\lesssim$3 Myr).

\subsection{Accretion parameters vs. stellar properties}

\begin{figure}[t]
\centering
    \includegraphics[width=0.48\textwidth]{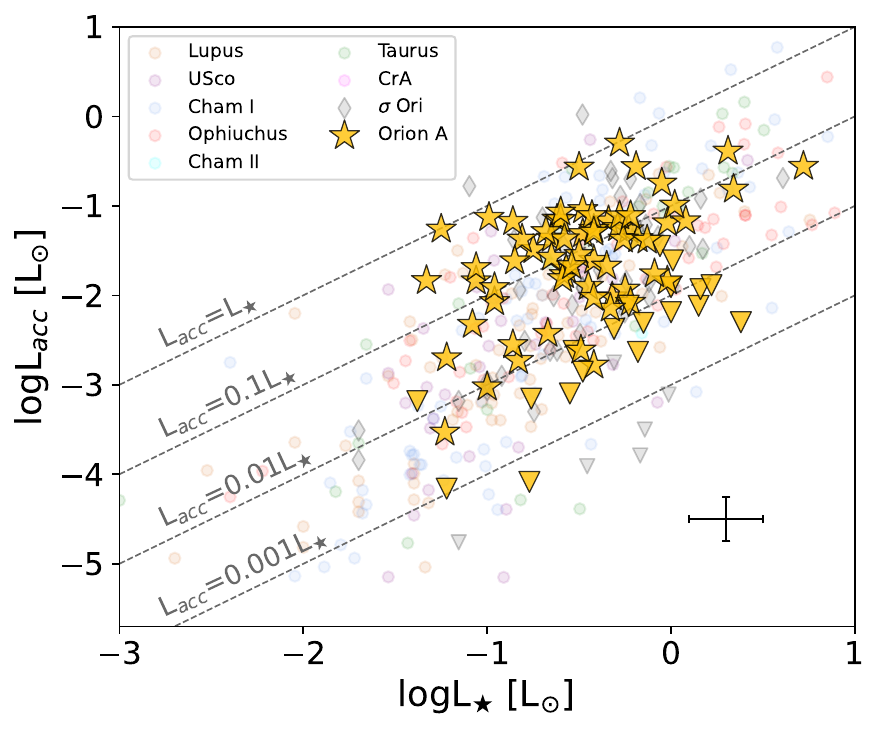}
\caption{Accretion luminosity as a function of stellar luminosity. The Orion A sample is shown as yellow stars with triangles indicating upper limits on \Lacc. The colored transparent circles are the sample compiled by \citet{Manara_2023} plus the $\sigma$ Ori sample from \citet{Mauco_2023}. The typical uncertainties for the accretion and stellar luminosities are shown ($\sim$0.25 and $\sim$0.2\,dex, respectively). The gray dashed lines mark the regions of constant \Lacc/\Lstar\ ratio, as labeled.}
\label{fig:Lacc_Lstar}     
\end{figure}

Our new Orion A spectra provide the largest dataset to date of YSOs in the Orion complex for which accretion properties have been consistently determined from modeling the UV excess. Here, we explore several accretion scaling relations, starting with the dependence of the accretion luminosity on the stellar luminosity. We observe in Fig.~\ref{fig:Lacc_Lstar} that most Orion A sources studied in this work ($\sim$80\%) show \Lacc$>$0.01\,\Lstar\ in line with the accretion luminosities retrieved from samples in other SFRs. From the remaining $\sim$20\% of the stars showing \Lacc$<$0.01\,\Lstar\, most are reported here as accretion upper limits (Sect.~\ref{subsec:acc_limits}). These stars still show IR excess emission according to Spitzer, indicating the presence of a disk despite no measurable accretion signature. This finding is consistent with the result that accretion ceases on shorter timescales than disk dissipation \citep[e.g.,][]{Fedele_2010,Briceno_2019,Delfini_2025}, so disk-bearing sources with undetectable accretion can already be present in young regions such as Orion A. The fraction of non-accretors should increase with region age, and indeed accretion upper limits evolves from $\sim$6\% in the younger Lupus SFR \citep{Alcala_2017} to $\sim$50\% in the older Upper Scorpius \citep{Empey_2026}. Our estimate falls between these two, though we note that the Orion A sample presented here does not reach a level of completeness comparable to the Lupus and Upper Scorpius surveys. We also stress that some of these accretion upper limits may correspond to low accretors rather than true non-accretors \citep[e.g.,][]{Thanathibodee_2022}. Nevertheless, the majority of our sample is accreting, with sources reaching \Lacc$\sim$\Lstar.

As for the other regions, we observe an increase of the accretion luminosity with increasing stellar luminosity. Given the brightness cut we performed in our sample selection, we do not cover the stellar luminosity range uniformly in order to test whether the \Lacc$-$\Lstar\ relation in Orion A is better described by a single or a broken power law as it has been tested in other SFRs (e.g., \citealt{Manara_2017a,Alcala_2017,Almendros-Abad_2024}). Most of our sources cluster around the luminosity regime where a potential break could occur (\Lstar$\sim$10$^{-0.5} L_{\odot}$) and we are particularly missing statistics at very low-luminosity or sub-stellar domain. We can, however, observe that within the luminosity range we sample, our sources populate the same region of the \Lacc$-$\Lstar\ parameter space as those in other nearby SFRs, recovering the full extend of the spread ($\gtrsim$2\,dex) despite Orion A being the most massive SFR among the ones analyzed.

\begin{figure*}[h!]
\centering
    \includegraphics[width=0.85\textwidth]{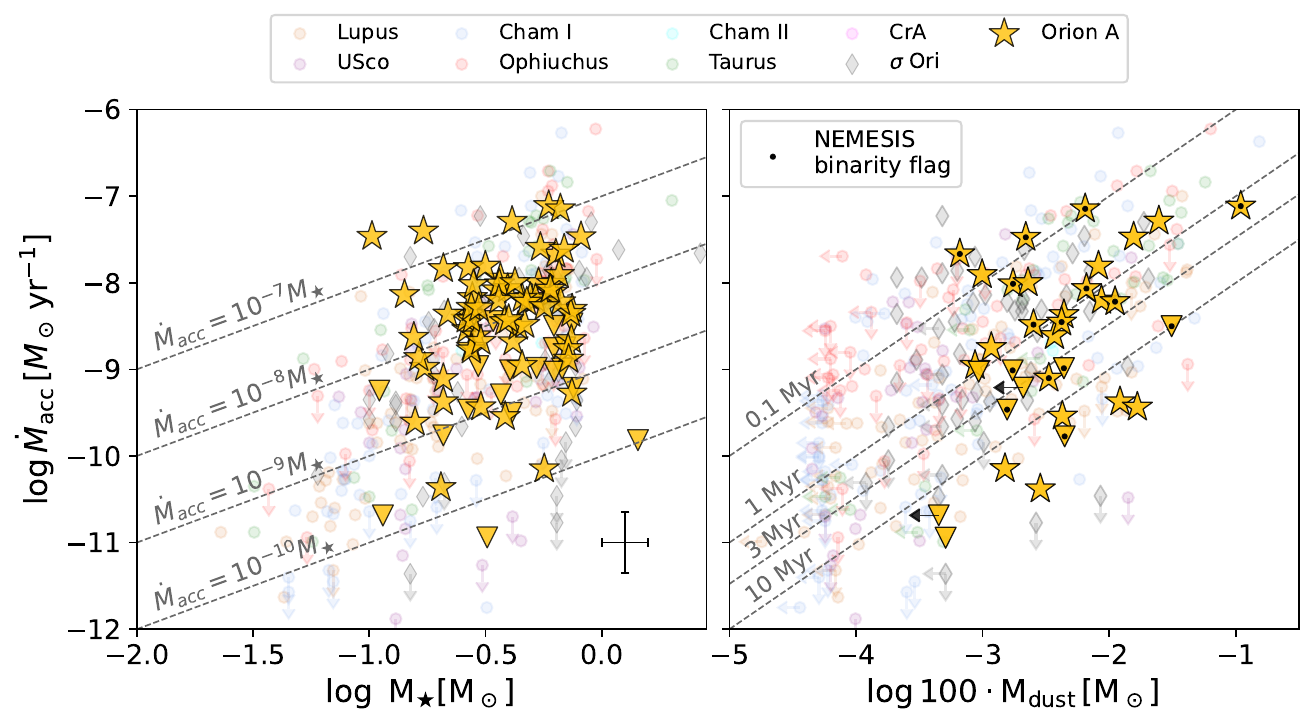}
\caption{Correlations between stellar, accretion and disk properties. The marker scheme is similar to that of Fig.~\ref{fig:Lacc_Lstar}. \textit{Left panel:} Mass accretion rate as a function of stellar mass. The triangles indicate the upper limits on \Macc. The error bars at the bottom right represent the typical uncertainties of 0.35 and 0.1\,dex for \Macc\ and \Mstar, respectively. The gray dashed lines mark the regions of constant \Macc/\Mstar\ ratio, as labeled. \textit{Right panel:} Mass accretion rate as a function of total disk mass. The disk masses were obtained considering a gas-to-dust ratio of 100. The triangles indicate the upper limits on \Macc, while the horizontal arrow correspond to the upper limit on \Mdisk. Probable binaries are pinpointed with a black dot. The dashed gray lines show disk lifetimes at relevant ages.}
\label{fig:Macc_Mstar_Mdisk}      
\end{figure*}

Similarly, as can be seen in the left panel of Fig.~\ref{fig:Macc_Mstar_Mdisk}, the distribution of the estimated \Macc\ as a function of the \Mstar\ is comparable to those found in other SFRs. These include the young and relatively isolated regions of Taurus and Lupus ($\sim$1$-$3 Myr), the more irradiated environment of $\sigma$ Ori ($\sim$3$-$5 Myr) and the older Upper Scorpius ($\sim$5$-$10 Myr). We do not attempt to constrain the power-law slope of the relation for Orion A sources since it is confined to the narrow stellar mass range $\sim$0.1$-$1.0 \Msun\ where other SFRs cover a broader range.
Within the same star-forming complex, we probe two sub-regions characterized by different stellar densities and irradiation environments (extended ONC and L1641/L1647 regions). We investigated whether these environmental differences imprint any systematic trend in the \Lacc$-$\Lstar\ and \Macc$-$\Mstar\ planes, finding that both populations occupy the same parameter space and show a comparable spread with no evident systematic differences (App.~\ref{app:regions_scaling_relations}). 

The higher accretors are sources \#18, \#43, \#53, \#66, and \#85 with estimated \Macc$\gtrsim$10$^{-7}$\Msun. Source \#85 is particularly interesting as it is one of the lowest mass sources in our sample (\Msun$\sim$0.1\,\Msun) and shows atypically high accretion rate. Its spectrum is dominated by a forest of emission lines across the UVB and VIS arms, whose nature and interpretation merit further future investigation.

Our work demonstrates that the behavior of \Macc$-$\Mstar\ relation hold even across the Orion A complex which harbors YSOs ($\sim$1$-$3 Myr) in a variety of local environments. This is consistent with the emerging picture of similar accretion–stellar scaling relations across SFRs of different environments and ages \citep[e.g.,][]{Testi_2022, Manara_2023, Almendros-Abad_2024, Delfini_2025}. However, more complete and homogeneous samples spanning a broader range of environments and ages remain crucial for testing this apparent universality and constraining disk evolution. In this context, new results in Upper Scorpius \citep{Empey_2026} suggest increasing scatter in the accretion–stellar relations at older ages, possibly indicating a weakening with time. Expanding to similar sample sizes ($>$100 disks) in regions such as the extended ONC or $\sigma$ Ori will be key, though currently both our sample and that of \citet{Mauco_2023} for $\sigma$ Ori show no significant differences with respect to less dense and externally irradiated environments such as Lupus and Taurus.

\subsection{Mass accretion rate vs. disk mass}\label{discussion:accretion_diskmass}

Under the assumption of viscous evolution of protoplanetary disks, a fundamental prediction is that \Macc\ should scale with the \Mdisk\ \citep{Hartmann_1998,Lodato_2017,Rosotti_2017,Mulders_2017}. 
Self-similar solutions predict \Mdisk$\sim$\Macc$\tau_{\nu}$, where $\tau_{\nu}$ is the viscous timescale at the outer disk radius \citep{Lynden-Bell_1974,Hartmann_1998}. As a consequence, in a viscously evolved disk, the ratio $t_{\rm disk}=$\Mdisk/\Macc\ should be comparable to the system age \citep{Jones_2012,Manara_2016b,Rosotti_2017}, representing the timescale over which the current disk mass would be depleted by accretion alone. This ratio is typically known as the disk lifetime \citep[e.g.,][]{Lodato_2017}.
Observational efforts have confirmed this \Macc$-$\Mdisk\ scaling (albeit with an intrinsic large scatter) by combining millimeter dust continuum measurements with accretion diagnostics from UV or optical spectroscopy \citep{Manara_2016b,Mulders_2017,Testi_2022,Manara_2023}, yet the Orion A cloud remains relatively unexplored in this context.

Dust mass estimates rely on assumptions about the gas-to-dust ratio (typically assumed to be 100), dust opacity, disk temperature, and optical depth. In particular, millimeter continuum emission is commonly used as a proxy for the total dust mass, under the assumption that the emission is optically thin and that a single representative dust temperature can characterize the disk (\citealt{Miotello_2023} and references therein). Despite these assumptions, millimeter-derived dust masses remain the most widely available and homogeneous constraint across large disk samples. 
With these caveats in mind, of the 84 sources in our sample for which reliable accretion rate estimates could be derived, 34 have reported disk dust mass measurements in the literature (App.~\ref{app:dust_masses} for more details). Assuming the canonical gas-to-dust ratio of 100, these can be converted into total disk masses.

In Fig.~\ref{fig:Macc_Mstar_Mdisk} (right panel), we compare our mass accretion rates estimates from VLT/X-Shooter spectra with the total disk masses obtained from ALMA continuum observations. As for the \Lacc$-$\Lstar\ and \Macc$-$\Mstar\ scaling relations, the distribution of our data on the \Macc$-$\Mdisk\ plane is similar to that observed in other SFRs. To the best of our knowledge, this represents the first exploration of the \Macc$-$\Mdisk\ relation in the Orion A molecular cloud that relies on homogeneously derived UV-based \Macc\ estimates.

As mentioned previously, under the viscous scenario, one would expect the disk lifetime (defined by the ratio \Mdisk/\Macc) to be comparable to the system's age. From Fig.~\ref{fig:Macc_Mstar_Mdisk} (right panel), we observe that $\sim$56\% of the sources (19 sources) show disk lifetimes consistent with the age of the region ($\sim$1$-$3 Myr), $\sim$18\% (6 sources) show $t_{\rm disk}$ younger than 1 Myr, and other $\sim$26\% similar or older than 10 Myr. Hence, the fraction of sources with disk lifetimes in agreement with the expected age of Orion A YSOs is similar to that of sources with disk lifetimes much younger or older with respect to $\sim$1$-$3 Myr.
The large spread in disk lifetimes demonstrates that viscous evolution alone is unlikely to account for the full observed distribution. In reality, different physical processes impact differently the \Macc$-$\Mdisk\ plane \citep[e.g.,][]{Rosotti_2017,Manara_2023}: sources with short disk lifetimes may be undergoing enhanced mass loss through external photoevaporation, which depletes the disk in an outside-in fashion therefore reducing \Mdisk\ \citep{Johnstone_1998, Rosotti_2017,Sellek_2020, Winter_2022,Allen_2025}, or be in multiple systems, where tidal truncation results in shorter disk lifetimes at a given stellar age \citep{Zagaria_2022}. 
Conversely, sources with longer inferred disk lifetimes may reflect the influence of internal photoevaporation \citep[e.g.,][]{Alexander_2014,Ercolano_Pascucci_2017} or planet formation \citep[e.g.,][]{Manara_2019}, both of which can suppress \Macc\ while leaving \Mdisk\ relatively intact, potentially giving rise to transition disk morphologies \citep{vanDerMarel_2023}. With this in mind, we refer to $t_{\rm disk}$ as inferred disk lifetime hereafter.

Focusing on the role of multiplicity in driving short inferred disk lifetimes, \citet{Zagaria_2022} found that multiple stellar systems in Lupus and Upper Scorpius account for a significant fraction of sources with high accretion rates relative to their disk mass compared to isolated stars. 
To test whether this holds in our sample, especially given the clustering of five sources around \Mdisk/\Macc$\sim$0.1 Myr, we performed a search for potential binaries in our sample by considering the comprehensive NEMESIS (New Evolutionary Model for Early stages of Stars with Intelligent Systems) catalogue of YSOs in the Orion star forming complex compiled by \citet{Roquette_2025}. We cross-matched our sources with the NEMESIS catalogue using a matching radius of 0.5'' and retained sources with the following multiplicity flags: \textit{A} (astrometric binary), \textit{E} (eclipsing binary), and \textit{S} (spectroscopic binary). We retrieved 15 sources (Table~\ref{tab:log_obs}\textcolor{blue}{.1}) and pinpoint those in the right panel of Fig.~\ref{fig:Macc_Mstar_Mdisk} with black dots along with sources \#34, \#62, and \#88, flagged as tight binaries by \citet{Flaherty_2025}. 

We observe that four out of six sources consistent with $t_{\rm disk}\sim$0.1 Myr are binaries according to the NEMESIS catalog, suggesting that multiplicity has a significant impact on the observed scatter in our sample. However, not all binaries show anomalously short inferred disk lifetimes, which is expected since the truncation effect depends on the binary configuration: theoretically, closer binaries should have more tidally truncated disks and thus shorter inferred disk lifetimes, while circumbinary or wide binary configurations should show less truncation. Furthermore, our analysis is limited to systems with specific multiplicity classifications, and the true multiplicity fraction in our sample is likely underestimated given the incompleteness of available multiplicity surveys. While elevated RUWE values could in principle flag additional unresolved companions, this indicator is generally unreliable for YSOs as disk emission and accretion variability can mimic binarity signatures \citep{Belokurov_2020, Fitton_2022}. A more complete multiplicity census of L1641 YSOs is therefore needed to firmly quantify its role in the observed spread.

\subsection{Environmental dependences}\label{subsubsec:environment}

Most of the sources in our sample are located outside the bright central region of the ONC, where the Trapezium Cluster sits. Our closest star (\#15) to the prominent UV source in the Trapezium $-$ $\theta^{1}$~Ori~C $-$ is located at 2.34' ($\sim$0.3\,pc). In the Trapezium, roughly within the projected distances of $\sim$0.3\,pc, external UV radiation from the massive stars strongly affect disks, as evidenced by truncated disk mass distributions and the concentration of proplyds (externally irradiated protoplanetary disks with a cometary shape and an offset bright ionization front) within this radius (e.g., \citealt{Storzer_Hollenbach_1999,Ricci_2008,Mann_Williams_2010,Mann_2014,Eisner_2018,Aru_2024a}). Proplyds have also been identified in the northern sub-region NGC~1977 around the B-type star 42~Ori \citep{Kim_2016_proplyds}, demonstrating that B-type stars can drive external photoevaporation and generate proplyds.
Even outside of the immediate vicinity of OB stars, disks in the extended ONC ($d>$0.3\,pc from $\theta^{1}$ Ori C) can still be moderately irradiated. The local FUV field strength is typically expressed in units of G$_0$, where 1~G$_0$ corresponds to the interstellar radiation field in the solar neighborhood ($1.6\times10^{-3}$ erg s$^{-1}$ cm$^{-2}$). Moderately irradiated disks typically experience FUV fields of $\sim$10$^{2}-$10$^{4}$\,G$_{0}$ \citep{Anania_2026} being these strong enough to influence disk evolution \citep{Anania_2025b}.
Furthermore, moving away from the Orion Nebula, sources in the southern part of the Orion A complex (L1641/L1647 regions) are also irradiated to a certain degree by distributed B- and A- type stars, albeit with lower FUV fields ($\lesssim$10$^{2}$\,G$_{0}$; \citealt{Anania_2026}). Even in this regime, there is evidence of decreasing median dust disk mass with increasing intermediate FUV radiation fields
\citep{vanTerwisga_2023,vanTerwisga_2023_corrigendum}. 

To quantify the irradiation conditions experienced by our targets, we computed the incident FUV fields using the method presented by \citet{Anania_2025}, which seek to reconstruct the local 3D stellar density distribution to estimate the separation between disk-bearing stars and nearby massive stars. The resulting FUV flux estimates span almost four orders of magnitude: $\sim$31\% of disks experience weak irradiation ($<$100\,G$_{0}$), $\sim$38\% lie in the intermediate 10$^{2}$–10$^{3}$\,G$_{0}$ regime, $\sim$26\% are exposed to strong fields of 10$^{3}$–10$^{4}$\,G$_{0}$, and $\sim$5\% (four sources: \#15, \#17, \#21, and \#26) reach ONC–core levels ($\gtrsim$10$^{4}$\,G$_{0}$). These values indicate that most of our sample experience non-negligible external FUV fields (i.e., intermediate values between 10$^{2}$–10$^{4}$\,G$_{0}$) caused by member and field OBA-type stars across Orion A \citep{Anania_2026}.

Given the broad FUV field range covered by our sample ($\sim$5 orders of magnitude), we examined whether the local FUV irradiation influences \Macc, finding no statistically significant correlation ($\rho=$0.05$\pm$0.12, slope$=$0.04$\pm$0.11). Normalizing by $M_{\star}^{2}$ to remove the primary stellar mass dependence \citep[e.g.,][]{Manara_2023,Winter_2024} results in a marginally larger but still insignificant correlation ($\rho=$0.18$\pm$0.12, slope$=$0.17$\pm$0.11), with a large intrinsic scatter of $\sim$0.8\,dex in both cases. These findings are consistent with results in $\sigma$ Ori \citep{Rigliaco_2012,Winter_2020,Mauco_2023}.

We next examined how the inferred disk lifetime depends on FUV strength, using the 34 sources in our sample with available \Mdisk\ measurements (App.~\ref{app:dust_masses}).
Earlier studies have found hints of this effect: \citet{Rosotti_2017} reported shorter inferred disk lifetimes in the ONC than in Lupus, attributed to the much stronger FUV field in the ONC, while \citet{Mauco_2023} found an initial hint of shorter $t_{\rm disk}$ at closer projected distances from the massive OB stars in $\sigma$ Ori. Building on these results, and while the dependence of dust disk mass on FUV irradiation has been explored across multiple regions \citep{vanTerwisga_2023,vanTerwisga_2023_corrigendum}, a similar analysis in terms of $t_{\rm disk}$ (which simultaneously encodes both the accretion and disk mass properties) has not yet been attempted over a wide FUV baseline. Given that the subset of our sample with available mm counterparts is dominated by sources in the L1641 region, it only reaches FUV fields $\lesssim$10$^{3}$\,G$_{0}$.
We therefore extended our comparison by including sources in the core of the ONC \citep{Manara_2012, Eisner_2016, Mann_2014, Rosotti_2017} and in Lupus \citep{Manara_2023}, which share a similar age range of $\sim$1$-$3 Myr with our sample, as well as $\sigma$ Ori, which despite being somewhat older ($\sim$3$-$5 Myr; e.g.,  \citealt{Oliveira_2004}) is currently the only region with both accretion rate \citep{Mauco_2023} and disk mass measurements \citep{Ansdell_2017,Mauco_2023} in the $\sim$10$^{2}$--10$^{4}$\,G$_{0}$ range. 
Together, these regions span nearly five orders of magnitude in FUV flux ($\sim$1 to $\gtrsim$10$^{5}$\,G$_{0}$), representing the widest FUV baseline assembled for this type of analysis to date. For consistency, FUV estimates for all regions are computed following \citet{Anania_2025}, and all mass accretion rate estimates retrieved were obtained from UV excess modeling, with the exception of ONC sources for which $U$-band and H$\alpha$ diagnostics were used \citep{Manara_2012}.

\begin{figure}[t]
\centering
    \includegraphics[width=0.49\textwidth]{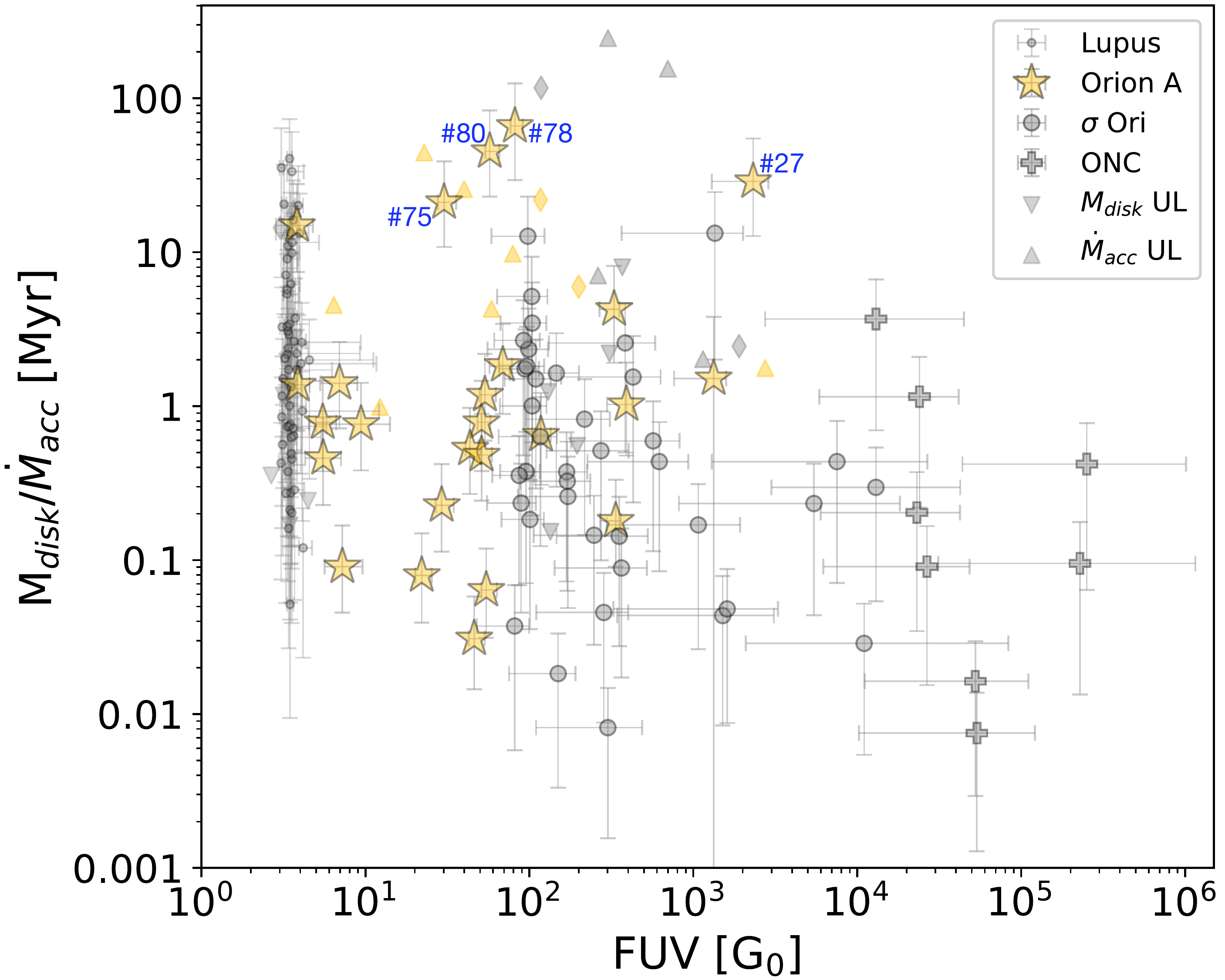}
\caption{Distribution of inferred disk lifetimes estimates as a function of FUV field strength for YSOs in our sample with ALMA observations (yellow markers), in Lupus (\citealt{Manara_2023}; dot markers), in $\sigma$ Orionis (\citealt{Mauco_2023}; circle markers), and in the ONC (\citealt{Rosotti_2017}; plus markers). Upward triangles indicate upper limits (UL) on \Macc, whereas downward triangles upper limits on \Mdisk.}
\label{fig:lifetime_UV}     
\end{figure}

Figure \ref{fig:lifetime_UV} hints for a tentative population-level decrease of inferred disk lifetimes at stronger FUV environments, but any potential dependence is currently dominated by large intrinsic scatter ($\gtrsim$2\,dex) and remains limited by the small number of sources with both \Mdisk\ and \Macc\ measurements at intermediate and high FUV fields. Nonetheless, the majority of sources experiencing FUV fields $>$10$^{3}$\,G$_{0}$ show $t_{\rm disk}\lesssim$1 Myr, whereas sources at lower FUV fields can reach higher values. Whether this reflects a gradual strengthening of the FUV dependence or a more abrupt transition (a ``switch'') above some threshold field strength cannot be determined with the present sample. 
We note that the ONC sample is the only one for which \Macc\ estimates are not spectroscopically derived, and is additionally biased towards the largest, most easily detectable disks. Nonetheless, even when not considering these sources, those at FUV $\gtrsim$10$^3$\,G$_0$ ($\sigma$ Ori) tend to show $t_{\rm disk} \lesssim$1 Myr, while sources at lower FUV fields display a broader range of values, in line with \citet{Mauco_2023} findings in the $\sigma$ Ori cluster itself.
Part of the scatter at low FUV fields is impacted by sources in our sample at the boundary between accretion and chromospheric emission (namely sources \#27, \#75, \#78, and \#80; Sect.~\ref{subsec:acc_limits}). These sources are considered as accreting in our analysis, though their ambiguous nature makes them natural outliers at the long inferred disk lifetime end.

A decrease in $t_{\rm disk}$ with FUV field strength would be consistent with external photoevaporation driving disk dispersal, in line with the lower dust disk masses in more strongly irradiated environments \citep[e.g.,][]{Mann_2014,Eisner_2018,vanTerwisga_2023,vanTerwisga_2023_corrigendum}. Establishing whether external photoevaporation is genuinely the dominant driver, however, remains challenging. \citet{Parker_2021}, for instance, caution that statistical limitations in current samples can reproduce such a hint, a possibility we cannot rule out here. 
The assumption of a static FUV environment adds a further layer of uncertainty: if massive OB stars have high proper motions, current projected separations may not accurately reflect the true irradiation history of individual disks. This could, however, work in favor of the comparison in Fig.~\ref{fig:lifetime_UV} since disks in $\sigma$ Ori, for instance, may be exposed to a FUV field that was activated only recently, following the proposed runaway nature of $\sigma$ Ori AB \citep{Coleman_2025b}. This would bring their effective time-integrated FUV exposure closer to that of the younger regions considered, partly offsetting the fact that $\sigma$ Ori is otherwise the oldest region in the comparison. Earlier works, however, found no such deviation \citep[e.g.,][]{Caballero_2007}, thus further observations are needed to confirm this scenario.
Notwithstanding the aforementioned caveats, the present analysis therefore represents the most comprehensive comparison currently possible with existing data.

\section{Conclusions}\label{sec:conclusions}

Observationally constraining accretion properties in YSOs in different environments and analyzing how these depend on both stellar and disk properties is key to 
better understand disk evolution and, ultimately, the timescales over which protoplanetary disks disperse. In this work, we present the largest VLT/X-Shooter study of accretion in Orion A. All of the sources targeted in this work have a ``Disk'' classification from infrared studies with Spitzer \citep{Megeath_2012}. Our sample is composed of 91 spectra of Class II YSOs (51 of which are newly obtained), from NGC 1977 in the north through the Orion Nebula and down to the L1641 and L1647 clouds. 
Here we list the main results of our study:

\begin{itemize}
    \item We self-consistently derived stellar and accretion properties for 84 PMS stars located in Orion A from broadband (UV to NIR) flux-calibrated spectroscopic data. The sample presented encompass YSOs with SpTs between K3 and M5 and \Macc\ $\sim$1$\times$10$^{-11}-$8$\times$10$^{-8}$ \Msun\,yr$^{-1}$. 

    \item After correcting for extinction and accretion, five sources remain sub-luminous by $\sim$1,dex compared to the remaining population (\#38, \#39, \#48, \#57, and \#75). Under the assumption of gray extinction, we used the [\ion{O}{i}] line to compute an obscuration factor and correct log\Lstar\ accordingly, bringing these sources into closer agreement with the rest of the population, though the size of the inferred correction varies considerably among them. 
    
    \item We analyzed the \Lacc$-$\Lstar\ and \Macc$-$\Mstar\ scaling relations in Orion A and compared them with those observed in other SFRs spanning a wide range of environments and ages. Despite these differences, all regions populate similar locus in the explored parameter spaces. By combining our \Macc\ measurements with dust mass estimates from ALMA for a subset of 34 sources, we also investigated, for the first time, the \Macc$-$\Mdisk\ correlation in Orion A, recovering a similar spread to that observed in other SFRs. 
    
    \item Four out of six sources clustered around $t_{\rm disk} \sim$0.1 Myr are identified as binaries, suggesting that multiplicity contributes to the observed scatter, although the remaining binaries recover the full spread of the \Macc$-$\Mdisk\ relation, indicating that it is not the dominant driver of the overall scatter. A more complete multiplicity census in Orion A is needed to firmly quantify its role.

    \item By implementing \citet{Anania_2025} method to estimate FUV fluxes at the position of our sample of 91 YSOs, we estimated that $\sim$31\% of disks experience weak irradiation ($<$100 G$_{\text{0}}$), $\sim$38\% lie in the intermediate 10$^{2}-$10$^{3}$ G$_{\text{0}}$ regime, $\sim$26\% are exposed to strong fields of 10$^{3}-$10$^{4}$ G$_{\text{0}}$, and $\sim$5\% (four sources: \#15, \#17, \#21, and \#26) reach ONC–core levels ($\gtrsim$10$^{4}$ G$_{\text{0}}$). These values indicate that most of our sample experience non-negligible external FUV fields;

    \item Despite spanning nearly five orders of magnitude in FUV field strength, we found no statistically significant correlation with \Macc, consistent with previous findings in $\sigma$ Ori \citep{Rigliaco_2012, Winter_2020, Mauco_2023}.

    \item We present the most comprehensive comparison to date of $t_{\rm disk}$ as a function of local FUV field strength, finding a tentative population-level decrease toward stronger FUV environments consistent with previous works \citep[e.g.,][]{Rosotti_2017,Mauco_2023}, though large intrinsic scatter ($\sim$2\,dex) and limited sample sizes, especially at intermediate and high FUV regimes, prevent firm conclusions. Improving the census of \Macc\ and $M_{dust}$ in these environments is therefore crucial to firmly assess the impact of FUV irradiation in disk evolution. 
\end{itemize}

The homogeneous \Macc\ measurements presented in this work, spanning a wide range of FUV field strengths, provide a solid foundation for future studies of disk evolution across environments. The characterization of accretion in young stars in the UCL/LCC regions of the Sco-Cen complex as part of ARADE will be presented in a forthcoming paper (Piscarreta et al. in prep).

\section*{Data availability}

Table \ref{tab:fitter_output} and the table with the \Lacc\ estimates from emission lines (Sect.~\ref{subsec:Lacc_lines}) are only available in electronic form at the CDS via anonymous ftp to \url{cdsarc.u-strasbg.fr} (130.79.128.5) or via \url{http://cdsweb.u-strasbg.fr/cgi-bin/qcat?J/A+A/}.
Additional data for this article are available at \url{https://doi.org/10.5281/zenodo.22212301}.

\begin{acknowledgements}
We thank the anonymous referee for their constructive and insightful feedback which improved the clarity of the manuscript.

The work presented is based on observations collected at the European Southern Observatory under ESO programme 108.2206, 114.276M, and 116.2934.

L.P. acknowledges the PhD fellowship of the International Max-Planck-Research School (IMPRS) funded by ESO. 

C.F.M. is funded by the European Union (ERC, WANDA, 101039452). Views and opinions expressed are however those of the author(s) only and do not necessarily reflect those of the European Union or the European Research Council Executive Agency. Neither the European Union nor the granting authority can be held responsible for them.

V.A-A acknowledges support from the INAF grants 1.05.12.05.03 and 1.05.24.07.02.

T.J. acknowledges the support from the MUNI Award in Science and Humanities (MUNI/I/1762/2023).

This work has made use of data from the European Space Agency (ESA) mission
{\it Gaia} (\url{https://www.cosmos.esa.int/gaia}), processed by the {\it Gaia} Data Processing and Analysis Consortium (DPAC, \url{https://www.cosmos.esa.int/web/gaia/dpac/consortium}). Funding for the DPAC has been provided by national institutions, in particular the institutions participating in the {\it Gaia} Multilateral Agreement.

This research has made use of data from the Herschel Gould Belt survey (HGBS) project (http://gouldbelt-herschel.cea.fr). The HGBS is a Herschel Key Programme jointly carried out by SPIRE Specialist Astronomy Group 3 (SAG 3), scientists of several institutes in the PACS Consortium (CEA Saclay, INAF-IFSI Rome and INAF-Arcetri, KU Leuven, MPIA Heidelberg), and scientists of the Herschel Science Center (HSC).

This research has made use of the SVO Filter Profile Service ``Carlos Rodrigo'', funded by MCIN/AEI/10.13039/501100011033/ through grant PID2023-146210NB-I00
\end{acknowledgements}

\bibliographystyle{aa}
\bibliography{XShoo_spec}

\begin{appendix} 

\onecolumn

\section{Additional tables}

Table \ref{tab:log_obs} summarizes the observing log of the different VLT/X-Shooter programs presented in this work along with estimated S/N in the different arms. We also include a note on whether sources are flagged as binaries in the NEMESIS catalog of Orion \citep{Roquette_2025} or studied by \citet{Flaherty_2025}.
Table~\ref{tab_app:gaia_info} summarizes the Gaia DR3 information.

\renewcommand{\arraystretch}{1.17}

\begin{longtable}{lccccccc}
\caption{Log of observations and S/N of the reduced VLT/X-Shooter spectra.}\\ 

\hline\hline
ID & Obs. date & RA & DEC & S/N$_{\rm UVB}$ & S/N$_{\rm VIS}$ & S/N$_{\rm NIR}$ & Note$^{a}$ \\
\hline
\endfirsthead

\caption{Continued.}\\
\hline\hline
ID & Obs. date & RA & DEC & S/N$_{UVB}$ & S/N$_{VIS}$ & S/N$_{NIR}$ & Note$^{a}$ \\
\hline
\hline
\endhead

\hline
\endfoot
\multicolumn{8}{c}{Prog. ID 108.2206, PI Beccari}\\
\hline
01 & 2021-10-11T07:22:02.2361 & 05:35:16.07 & $-$05:20:36.3 & 8.7 & 22.7 & 28.2 & \\ 
02 & 2021-10-13T05:42:04.8883 & 05:34:52.17 & $-$05:22:31.9 & 6.6 & 18.8 & 24.5 & \\ 
03 & 2021-10-22T05:34:42.2531 & 05:35:57.65 & $-$05:57:18.4 & 10.1 & 29.2 & 32.3 & \\ 
04 & 2021-10-22T06:41:40.4609 & 05:35:08.02 & $-$05:32:44.3 & 7.8 & 29.8 & 25.5 &  \\
05 & 2021-10-22T07:17:34.0847 & 05:34:26.16& $-$05:26:30.4 & 3.1 & 29.7 & 40.8 & \\
06 & 2021-10-22T07:52:32.9067 & 05:35:03.57 & $-$05:29:26.3 & 5.4 & 18.4 & 31.5 \\ 
07 & 2021-10-27T06:47:36.4625 & 05:35:31.50 & $-$05:05:01.7 & 10.2 & 15.8 & 36.2 & \\ 
08 & 2021-11-06T06:40:30.3567 & 05:35:25.23 & $-$05:15:35.8 & 4.1 & 22.5 & 32.4 &  \\ 
09 & 2021-11-11T07:29:31.7526 & 05:35:28.10 & $-$05:29:33.5 & 2.9 & 10.3 & 94.9 &  \\ 
10 & 2021-11-17T04:19:06.3926 & 05:34:13.20 & $-$05:33:53.5 & 20.8 & 23.3 & 31.9 & \\ 
11 & 2021-11-17T05:23:28.8478 &  05:35:32.22 & $-$05:44:26.6 & 8.7 & 14.8 & 27.7 & \\ 
12 & 2021-11-17T06:26:32.9277 &  05:34:45.20 & $-$05:10:47.6 & 7.3 & 21.8 & 33.9 & \\ 
13 & 2021-11-18T03:25:52.2864 &  05:34:50.45 & $-$05:20:20.4 & 9.4 & 28.5 & 29.7 & \\ 
14 & 2021-11-18T04:00:54.2508 &  05:35:41.32 & $-$05:27:50.3 & 34.7 & 25.1 & 50.3 & \\
15 & 2021-11-18T05:04:12.3192 &  05:35:18.96 & $-$05:21:07.8 & 2.0 & 17.2 & 29.5 &  \\
16 & 2021-11-18T06:06:42.6428 &  05:35:30.48 & $-$05:28:30.6 & 9.8 & 23.5 & 27.4 &  \\
17 & 2021-11-18T07:09:48.6522 &  05:35:29.34 & $-$05:25:46.3 & 5.9 & 14.5 & 27.1 &  \\
18 & 2021-11-19T06:03:55.4934 &  05:35:36.36 & $-$05:31:37.8 & 9.8 & 20.3 & 56.9 &  \\
19 & 2021-11-19T07:06:26.7590 &  05:34:56.51 & $-$05:27:51.0 & 4.5 & 19.4 & 24.8 &  \\
20 & 2021-11-20T03:17:55.1371 &  05:35:22.62 & $-$05:14:11.2 & 2.3 & 17.1 & 27.5 &  \\
21 & 2021-11-21T03:33:24.6080 &  05:35:23.66 & $-$05:26:27.1 & 3.8 & 20.4 & 50.8 &  \\
22 & 2021-11-21T04:37:26.1739 &  05:34:42.74 & $-$05:28:37.5 & 5.2 & 12.7 & 27.1 &  \\
23 & 2021-11-28T06:01:43.7267 &  05:35:17.35 & $-$05:42:14.6 & 10.9 & 15.7 & 33.3 & \\
24 & 2021-12-03T05:39:50.2152 &  05:34:52.92 & $-$05:28:59.0 & 12.9 & 20.1 & 38.9 & \\
25 & 2021-12-06T02:53:16.9013 &  05:35:05.77 & $-$05:33:55.8 & 2.3 & 17.0 & 32.8 &  \\
26 & 2021-12-14T02:10:35.7437 &  05:35:24.46 & $-$05:26:31.5 & 2.3 & 16.4 & 36.3 &  \\
27 & 2021-12-18T03:04:36.5588 &  05:35:19.29 & $-$05:16:44.7 & 1.4 & 12.9 & 41.0 &  \\
28 & 2021-12-18T04:07:10.9360 &  05:35:05.93 & $-$05:27:11.1 & 0.5 & 15.1 & 33.5 &  \\
29 & 2021-12-18T06:10:36.0862 &  05:35:06.60 & $-$05:26:51.0 & 1.6 & 11.9 & 34.3 &  \\
30 & 2022-01-11T03:07:59.6364 &  05:35:25.50 & $-$05:45:44.8 & 9.8 & 30.1 & 27.1 &  \\
31 & 2022-01-02T04:00:27.0481 &  05:34:49.07 & $-$05:26:26.6 & 5.1 & 18.0 & 44.2 &  \\
32 & 2022-01-04T03:24:01.3798 &  05:35:06.66 & $-$05:25:02.9 & 0.7 & 7.7 & 30.6 &  \\ 
33 & 2022-01-10T04:41:42.5439 &  05:34:49.58 & $-$05:04:59.5 & 4.7 & 12.8 & 37.7 & \\ 
\hline
\multicolumn{8}{c}{Prog. ID 114.276M, PI Piscarreta}\\
\hline
34 & 2024-10-22T05:57:41.8493 &  05:34:59.01 & $-$05:44:29.5 & 8.0 & 20.2 & 28.4 & TB \\ 
35 & 2024-12-05T01:07:00.9893 &  05:35:14.04 & $-$05:52:09.0 & 7.4 & 25.6 & 21.5 &  \\ 
36 & 2025-01-19T01:15:56.1983 &  05:35:25.52 & $-$04:51:20.7 & 5.2 & 17.2 & 35.0 &  \\ 
37 & 2024-12-28T03:08:31.7252 &  05:35:30.90 & $-$04:55:18.0 & 24.9 & 28.7 & 29.0 & S \\ 
38 & 2024-12-27T04:22:12.8600 & 05:32:49.93  & $-$06:10:45.7 & 6.4 & 26.7 & 51.7 &  \\ 
39 & 2024-12-06T00:47:51.3829 & 05:33:20.99  & $-$06:22:33.3 & 8.5 & 31.1 & 45.2 &  \\ 
40 & 2025-01-19T02:17:25.4789 & 05:34:12.89  & $-$05:28:48.2 & 8.1 & 24.8 & 19.3 &  \\ 
41 & 2025-01-11T02:34:33.4493 &  05:40:45.12 &  $-$07:22:25.1 & 28.8 & 78.6 & 60.0 & \\ 
42 & 2025-01-18T04:22:36.2669 &  05:36:40.76 &  $-$06:11:08.3 & 4.4 & 24.4 & 30.8 & S  \\ 
43 & 2025-01-21T06:13:34.3902 &  05:39:22.34 &  $-$07:26:44.5 & 2.7 & 28.8 & 43.7 & A  \\ 
44 & 2025-02-05T03:03:09.7911 & 05:33:01.76  & $-$04:49:18.5 & 4.4 & 24.7 & 95.4 &  \\ 
45 & 2025-02-16T01:17:13.2978 &  05:35:42.47 &  $-$05:27:33.1 & 8.7 & 24.1 & 29.8 & \\ 
46 & 2025-02-19T01:32:00.0559 &  05:35:24.49 &  $-$06:01:46.4 & 23.4 & 19.7 & 48.8 & \\
47 & 2025-02-19T02:33:51.4648 &  05:35:02.98 & $-$04:48:32.8 & 18.9 & 16.2 & 35.5 & \\ 
48 & 2025-02-19T03:35:28.2624 &  05:35:08.70 & $-$04:46:52.5 & 3.9 & 26.4 & 58.8 &  \\ 
49 & 2025-02-23T02:58:52.3088 &  05:34:48.45 & $-$04:50:51.4 & 9.7 & 26.7 & 39.4 &  \\ 
50 & 2025-02-24T00:48:51.1719 &  05:35:28.69 & $-$04:48:16.4 & 9.7 & 25.8 & 29.0 &  \\ 
51 & 2025-02-24T01:53:52.9545 &  05:36:21.82 & $-$06:26:02.2 & 11.7 & 32.1 & 42.3 & S \\ 
52 & 2025-02-24T02:55:30.8902 &  05:37:15.50 & $-$06:59:03.5 & 8.7 & 26.8 & 31.5 &  \\
53 & 2025-02-24T03:57:20.7048 &  05:34:54.05 & $-$06:23:50.7 & 28.1 & 29.1 & 37.3 & \\ 
54 & 2025-02-25T03:28:39.1431 &  05:35:36.01 & $-$05:12:25.3 & 0.0 & 31.6 & 71.1 & \\ 
55 & 2025-02-26T01:04:30.3859 &  05:42:40.81 & $-$08:40:08.7 & 6.4 & 24.1 & 33.8 & \\ 
56 & 2025-02-26T02:06:17.7229 &  05:42:59.13 & $-$08:09:23.6 & 3.1 & 20.0 & 33.3 & A \\
57 & 2025-02-26T03:07:43.0307 & 05:33:59.24  & $-$05:46:23.2 & 2.7 & 14.7 & 25.5 &  \\ 
58 & 2025-02-27T03:10:06.5149 &  05:36:26.13 & $-$06:08:03.7 & 4.8 & 21.7 & 27.2 & S \\
59 & 2025-03-03T02:44:15.0099 &  05:35:18.39 & $-$04:53:23.6 & 1.7 & 23.2 & 51.8 & \\ 
60 & 2025-03-05T02:43:52.0847 &  05:35:18.52 & $-$05:13:38.3 & 0.0 & 0.6 & 50.0 & \\ 
61 & 2025-03-07T00:37:35.5661 & 05:34:27.65  & $-$04:57:05.1 & 11.7 & 28.4 & 62.7 & \\ 
62 & 2025-03-07T01:42:07.3417 &  05:36:06.61 & $-$05:41:54.3 & 6.3 & 22.4 & 48.9 & TB \\ 
63 & 2025-03-07T02:44:22.1706 &  05:35:15.66 & $-$06:01:27.9 & 0.5 & 8.7 & 29.9 &  \\
64 & 2025-03-08T01:13:56.2389 &  05:35:27.92 & $-$06:14:15.1 & 11.5 & 38.5 & 38.0 & S \\
65 & 2025-03-08T02:16:33.3867 &  05:41:28.20 & $-$09:23:40.5 & 1.0 & 14.0 & 26.3 &  \\ 
66 & 2025-03-08T03:18:06.3240 &  05:38:52.38 & $-$07:21:09.3 & 3.0 & 27.8 & 39.8 & A  \\
67 & 2025-03-09T02:59:39.1711 &  05:36:15.07 & $-$06:17:36.9 & 6.4 & 23.1 & 25.5 & E  \\
68 & 2025-03-13T00:34:20.9592 &  05:35:43.55 & $-$05:05:41.4 & 16.7 & 22.5 & 31.7 & \\
69 & 2025-03-13T01:07:24.7874 &  05:35:36.44 & $-$05:34:11.2 & 12.9 & 23.6 & 27.3 & \\
70 & 2025-03-13T01:40:22.0134 &  05:35:25.38 & $-$05:53:21.5 & 3.0 & 18.5 & 32.3 &  \\
71 & 2025-03-13T02:13:22.6715 & 05:34:29.27  & $-$05:14:39.7 & 5.0 & 21.3 & 24.7 &  \\
72 & 2025-03-13T02:46:16.5149 &  05:35:11.89 & $-$05:31:55.4 & 7.1 & 25.5 & 27.4 & \\ 
73 & 2025-03-20T01:34:59.8517 &  05:34:47.64 & $-$04:50:01.3 & 1.8 & 18.0 & 22.6 & \\ 
74 & 2025-03-28T01:48:27.4009 &  05:35:25.68 & $-$05:30:21.1 & 9.8 & 24.6 & 33.9 & \\ 
75 & 2025-03-30T01:21:36.9911 &  05:37:56.99 & $-$06:36:33.3 & 2.8 & 19.1 & 20.0 & \\ 
76 & 2025-03-31T00:25:06.4320 & 05:32:43.42  & $-$05:35:57.3 & 7.8 & 19.2 & 28.4 &  \\
77 & 2025-03-31T01:26:52.6686 & 05:34:02.18  & $-$05:36:19.6 & 7.3 & 26.9 & 34.7 &  \\
\hline
\multicolumn{8}{c}{Prog. ID 116.2934, PI Piscarreta}\\
\hline
78 & 2026-02-15T00:51:00.5687 &  05:34:34.69 & $-$06:08:27.8 & 4.1 & 11.3 & 43.8 &  \\ 
79 & 2026-02-15T03:54:38.4694 &  05:35:25.12 & $-$06:47:56.5 & 37.0 & 37.3 & 47.0 & \\  
80 & 2026-02-17T02:31:56.1237 &  05:34:47.66 & $-$06:19:40.1 & 7.1 & 20.6 & 34.7 &  \\ 
81 & 2026-02-17T03:36:05.8699 &  05:35:41.03 & $-$06:22:45.4 & 10.0 & 23.5 & 30.0 & S \\ 
82 & 2026-02-27T01:48:32.8054 &  05:41:33.22 & $-$07:55:02.3 & 16.2 & 27.3 & 31.6 & S \\ 
83 & 2026-03-01T02:38:52.2620 &  05:35:21.88 & $-$05:07:01.8 & 5.4 & 24.4 & 36.6 & \\
84 & 2026-03-25T01:00:00.9680 & 05:40:48.52 & $-$09:23:46.8 & 4.2 & 24.4 & 23.5 & \\ 
85 & 2026-03-29T00:43:46.0373 & 05:35:47.65 & $-$06:21:36.1 & 2.9 & 18.0 & 47.7 & S \\ 
86 & 2026-04-01T00:35:47.7037 & 05:36:37.04 & $-$05:04:41.1 & 6.2 & 17.4 & 24.3 & S \\ 
87 & 2026-04-04T23:52:55.9731 & 05:36:58.98 & $-$06:29:04.9 & 7.2 & 12.2 & 27.5 & S \\ 
88 & 2026-04-08T00:01:09.4953 & 05:35:03.92 & $-$05:29:03.4 & 8.8 & 29.2 & 37.3 & TB \\ 
89 & 2026-04-08T23:54:01.7162 & 05:34:57.16 & $-$07:01:47.4 & 6.7 & 15.6 & 31.4 & \\ 
90 & 2026-04-09T23:56:09.0712 & 05:38:26.04 & $-$07:37:59.1 & 10.3 & 14.9 & 26.2 & A \\
91 & 2026-04-15T23:51:13.2019 & 05:35:39.77 & $-$04:40:24.3 & 4.5 & 24.4 & 28.9 & 
\label{tab:log_obs}
\end{longtable}
\tablefoot{S/N computed for UVB, VIS, and NIR in the following wavelength ranges, respectively: [399, 402], [698, 702], and [1648,1652]\,nm. \\$^{a}$ Sources in the tight binary catalog of \citet{Flaherty_2025} (TB) or with a binarity flag from the NEMESIS catalog \citep{Roquette_2025} (A - astrometric binary; E - eclipsing binary; S - spectroscopic binary).}

\newpage

\renewcommand{\arraystretch}{1.2}
\begin{longtable}{lclcccc}
\caption{Gaia DR3 astrometry.}\\ 

\hline\hline
ID & Gaia DR3 & Name & $\varpi$ [mas] & $\mu^{*}_{\alpha}$ [mas\,yr$^{-1}$] & $\mu_{\delta}$ [mas\,yr$^{-1}$] & RUWE  \\
\hline
\endfirsthead

\caption{Continued.}\\
\hline\hline
ID & GaiaDR3 & Simbad Name & $\varpi$ [mas] & $\mu^{*}_{\alpha}$ [mas\,yr$^{-1}$] & $\mu_{\delta}$ [mas\,yr$^{-1}$] & RUWE \\
\hline
\endhead

\hline
\endfoot
01 & 3017365914449685888 &  V2280 Ori & 2.88$\pm$0.16 & 2.14$\pm$0.15 & 3.83$\pm$0.12 & 2.69 \\
02$^{*}$ & 3017364887967710976 &  V1458 Ori & 2.82$\pm$0.73 & 2.67$\pm$0.64 & 3.28$\pm$0.52 & 9.97 \\
03 & 3017186251673566208 &  AU Ori & 2.60$\pm$0.03 & 1.21$\pm$0.02 & $-$0.71$\pm$0.02 & 1.10 \\
04 & 3017265244711916672 &  LN Ori & 2.54$\pm$0.05 & 0.41$\pm$0.04 & $-$0.09$\pm$0.03 & 1.08 \\
05 & 3017270291298102528 &  V1956 Ori & 2.61$\pm$0.04 & 1.48$\pm$0.04 & $-$1.28$\pm$0.03 & 0.95 \\
06 & 3017266275504061056 &  V405 Ori & 2.15$\pm$0.09 & 3.63$\pm$0.09 & $-$0.01$\pm$0.06 & 1.15\\
07$^{*}$ & 3209530830107709696 &  V422 Ori & 2.66$\pm$0.52 & 3.86$\pm$0.51 & 4.93$\pm$0.41 & 21.34 \\
08 & 3017367636742162176 &  V2434 Ori & 2.56$\pm$0.04 & 3.54$\pm$0.04 & 1.43$\pm$0.04 & 1.08 \\
09 & 3017359184248230272 &  V421 Ori & 2.55$\pm$0.05 & 1.45$\pm$0.05 & $-$0.26$\pm$0.04 & 0.99 \\
10 & 3017266962698857216 &  IM Ori & 2.66$\pm$0.03 & 0.19$\pm$0.03 & 1.05$\pm$0.02 & 1.04 \\
11$^{*}$ & 3017245655363965696 &  V2511 Ori & 2.92$\pm$0.66 & $-$6.63$\pm$0.62 & 7.16$\pm$0.54 & 26.64 \\
12 & 3209525263830030336 &  XZ Ori & 2.54$\pm$0.04 & 0.96$\pm$0.03 & $-$2.07$\pm$0.03 & 1.28\\
13 & 3209518598040871296 &  V2057 Ori & 2.56$\pm$0.04 & 2.12$\pm$0.04 & 0.48$\pm$0.03 & 1.09\\
14 & 3017359665283930752 &  V1401 Ori & 2.83$\pm$0.08 & 1.47$\pm$0.07 & 1.10$\pm$0.06 & 1.69\\
15 & 3017365850035788800 &  V2343 Ori & 2.44$\pm$0.06 & 0.80$\pm$0.06 & 0.64$\pm$0.05 & 0.78\\
16 & 3017359596564784128 &  V1292 Ori & 2.49$\pm$0.05 & 0.25$\pm$0.04 & $-$0.10$\pm$0.04 & 1.00 \\
17 & 3017360764795569792 & 2MASS J05352934-0525462 & 2.27$\pm$0.10 & 1.55$\pm$0.09 & 1.12$\pm$0.07 & 1.56 \\
18 & 3017347089620157440 &  V2551 Ori & 2.48$\pm$0.06 & 2.07$\pm$0.05 & 0.33$\pm$0.04 & 1.03 \\
19 & 3017269505319467520 &  V2080 Ori & 2.49$\pm$0.05 & 1.04$\pm$0.05 & $-$0.80$\pm$0.04 & 0.93 \\
20 & 3017367739821367552 & 2MASS J05352262-0514112 & 2.49$\pm$0.06 & 1.14$\pm$0.05 & 1.70$\pm$0.05 & 1.04 \\
21 & 3017360627356915328 &  V2411 Ori & 2.41$\pm$0.06 & 2.59$\pm$0.05 & $-$0.93$\pm$0.04 & 0.96 \\
22 & 3017269264801309568 &  V2012 Ori & 2.76$\pm$0.28 & 3.86$\pm$0.26 & 2.08$\pm$0.19 & 1.88 \\
23 & 3017246656093604736 &  V411 Ori & 2.51$\pm$0.04 & 1.81$\pm$0.04 & $-$1.16$\pm$0.03 & 1.19 \\
24 & 3017269367880519936 &  V1461 Ori & 2.51$\pm$0.07 & $-$0.76$\pm$0.06 & 0.79$\pm$0.05 & 1.05 \\
25 & 3017264626236630656 &  V408 Ori & 2.57$\pm$0.07 & 1.27$\pm$0.06 & 0.29$\pm$0.05 & 0.98 \\
26 & 3017360627356913280 &  V1283 Ori & 2.50$\pm$0.07 & 1.52$\pm$0.07 & $-$1.60$\pm$0.06 & 0.73 \\
27 & 3017367499303217280 &  V2347 Ori & 2.52$\pm$0.06 & 0.78$\pm$0.05 & 0.85$\pm$0.05 & 1.05 \\
28 & 3017360421198666624 &  V2162 Ori & 2.49$\pm$0.10 & 1.65$\pm$0.09 & $-$1.16$\pm$0.07 & 0.94 \\
29 & 3017360455558400896 & JW  296 & 2.47$\pm$0.08 & 2.52$\pm$0.07 & $-$0.85$\pm$0.05 & 0.88 \\
30 & 3017245144265120384 &  NR Ori & 2.21$\pm$0.07 & 2.07$\pm$0.05 & 0.18$\pm$0.05 & 1.61 \\
31 & 3017269848916851840 &  V2044 Ori & 2.55$\pm$0.08 & 2.31$\pm$0.07 & 1.03$\pm$0.05 & 1.01 \\
32 & 3017363582294575744 &  V2167 Ori & 2.35$\pm$0.16 & 1.03$\pm$0.15 & 0.61$\pm$0.10 & 0.89 \\
33 & 3209533029130884864 &  V2046 Ori & 2.51$\pm$0.05 & 1.53$\pm$0.04 & $-$1.04$\pm$0.04 & 0.99 \\
34 & 3017249095634906496 & CXORRS J053458.9-054429 & 2.50$\pm$0.04 & 0.61$\pm$0.04 & 0.81$\pm$0.03 & 1.65 \\
35 & 3017191646152505088 &  BZ Ori & 2.59$\pm$0.02 & 1.44$\pm$0.02 & 0.60$\pm$0.01 & 1.08 \\
36 & 3209572096153320192 &  V2438 Ori & 2.63$\pm$0.05 & 0.42$\pm$0.04 & $-$1.24$\pm$0.04 & 1.04 \\
37 & 3209559486129350912 &  AL Ori & 2.53$\pm$0.02 & 1.91$\pm$0.02 & $-$1.18$\pm$0.01 & 1.05 \\
38 & 3017205317032619008 & Haro 4-287 & 2.63$\pm$0.06 & 1.57$\pm$0.05 & 0.39$\pm$0.04 & 0.99 \\
39 & 3016776683592103552 &  V542 Ori & 2.54$\pm$0.05 & 1.28$\pm$0.05 & $-$0.01$\pm$0.04 & 1.21 \\
40 & 3209422253332724224 & Parenago  1382 & 2.61$\pm$0.02 & 0.48$\pm$0.02 & $-$0.12$\pm$0.02 & 1.12 \\
41 & 3016033100196424192 &  V902 Ori & 2.31$\pm$0.03 & 0.26$\pm$0.03 & $-$0.13$\pm$0.02 & 1.42 \\
42 & 3016980093241655936 &  V2703 Ori & 2.40$\pm$0.05 & 2.02$\pm$0.04 & $-$0.95$\pm$0.04 & 1.08 \\
43 & 3016028191049876096 & Haro 4-255 & 2.55$\pm$0.03 & 0.75$\pm$0.03 & $-$0.38$\pm$0.02 & 1.47 \\
44 & 3209608723634530944 &  HS Ori & 2.60$\pm$0.03 & 0.70$\pm$0.03 & 0.21$\pm$0.03 & 1.03 \\
45 & 3017359699643660160 &  OQ Ori & 2.56$\pm$0.03 & 1.14$\pm$0.03 & 1.02$\pm$0.02 & 1.19 \\
46 & 3017184808564584192 &  NS Ori & 2.67$\pm$0.04 & 1.12$\pm$0.04 & 0.92$\pm$0.03 & 1.14 \\
47 & 3209576013163656064 &  V2127 Ori & 2.52$\pm$0.05 & 2.02$\pm$0.04 & $-$2.05$\pm$0.04 & 1.02\\
48 & 3209576459840251776 &  V555 Ori & 2.54$\pm$0.06 & 1.14$\pm$0.05 & $-$0.24$\pm$0.04 & 1.26 \\
49 & 3209552339303795072 & Haro 4-44 & 2.51$\pm$0.05 & 0.92$\pm$0.05 & $-$0.72$\pm$0.04 & 1.02 \\
50 & 3209573092585717504 &  V418 Ori & 2.70$\pm$0.07 & 1.66$\pm$0.07 & $-$0.60$\pm$0.05 & 1.50 \\
51 & 3016948997678802048 &  V582 Ori & 2.58$\pm$0.11 & 1.08$\pm$0.09 & $-$0.38$\pm$0.08 & 1.50 \\
52 & 3016855092513584256 &  V587 Ori & 2.52$\pm$0.03 & 0.58$\pm$0.03 & 0.26$\pm$0.02 & 1.06 \\
53 & 3017134471547930624 &  BW Ori & 2.56$\pm$0.02 & 0.47$\pm$0.02 & 0.20$\pm$0.02 & 1.28 \\
54 & 3017373714110400128 &  NY Ori & 2.48$\pm$0.03 & 1.70$\pm$0.03 & $-$1.48$\pm$0.02 & 1.73 \\
55 & 3012331358078862720 & 2MASS J05424081-0840086 & 2.33$\pm$0.05 & 0.18$\pm$0.05 & $-$0.62$\pm$0.04 & 1.04 \\
56 & 3015730807514906880 & 2MASS J05425913-0809236 & 2.30$\pm$0.04 & 0.00$\pm$0.04 & $-$0.12$\pm$0.04 & 1.00 \\
57 & 3017253871638555520 &  V1913 Ori & 2.37$\pm$0.10 & $-$0.15$\pm$0.10 & 0.80$\pm$0.08 & 1.18 \\
58 & 3017168178451197952 &  V2679 Ori & 2.59$\pm$0.02 & 1.53$\pm$0.02 & 0.45$\pm$0.02 & 1.14\\
59 & 3209571580757261952 &  V2329 Ori & 2.57$\pm$0.09 & 1.34$\pm$0.08 & $-$1.22$\pm$0.06 & 1.01 \\
60 & 3209521415539408384 &  V2331 Ori & 2.44$\pm$0.08 & 0.51$\pm$0.07 & $-$0.79$\pm$0.06 & 0.97 \\
61 & 3209549831042932224 &  IR Ori & 2.46$\pm$0.05 & 1.77$\pm$0.04 & $-$0.07$\pm$0.04 & 1.06 \\
62 & 3017200545324683520 & 2MASS J05360660-0541543 & 2.54$\pm$0.09 & 1.82$\pm$0.08 & $-$0.49$\pm$0.06 & 1.16 \\
63 & 3017185049082755072 & [MGM2012] 1076 & 2.45$\pm$0.10 & 1.37$\pm$0.10 & 0.54$\pm$0.08 & 1.03\\
64 & 3017145913340772224 & Haro 4-364 & 2.46$\pm$0.03 & 1.79$\pm$0.03 & $-$0.10$\pm$0.02 & 1.16 \\
65 & 3012131590560373376 & 2MASS J05412819-0923404 & 2.32$\pm$0.07 & 0.85$\pm$0.05 & $-$1.25$\pm$0.05 & 1.07 \\
66 & 3016076878799144192 & Haro 4-254 & 2.50$\pm$0.02 & 1.56$\pm$0.02 & $-$0.86$\pm$0.02 & 1.34\\
67 & 3016977546325141376 &  BB Ori & 2.52$\pm$0.01 & 0.63$\pm$0.01 & $-$0.25$\pm$0.01 & 1.21 \\
68 & 3209529936754510208 &  AO Ori & 2.47$\pm$0.03 & 0.43$\pm$0.03 & 0.83$\pm$0.02 & 1.22 \\
69 & 3017346127547506048 &  V568 Ori & 2.55$\pm$0.03 & 1.81$\pm$0.03 & 0.30$\pm$0.02 & 1.19 \\
70 & 3017191336914852480 &  V1534 Ori & 2.51$\pm$0.03 & 1.15$\pm$0.03 & 0.85$\pm$0.02 & 1.10 \\
71 & 3209523442763914496 &  V1432 Ori & 2.50$\pm$0.06 & 1.12$\pm$0.05 & $-$0.76$\pm$0.04 & 4.60 \\
72 & 3017265279071650688 & Brun 556 & 2.52$\pm$0.02 & 2.60$\pm$0.02 & 1.53$\pm$0.01 & 1.18 \\
73 & 3209552442383007744 &  V2028 Ori & 2.61$\pm$0.07 & 0.59$\pm$0.06 & $-$1.06$\pm$0.05 & 1.01 \\
74 & 3017359012449385728 &  V417 Ori & 2.61$\pm$0.05 & 1.78$\pm$0.04 & 0.90$\pm$0.03 & 1.75 \\
75 & 3016906219804253312 & 2MASS J05375699-0636333 & 2.58$\pm$0.10 & 0.01$\pm$0.08 & $-$0.21$\pm$0.07 & 1.14 \\
76 & 3209393940908174080 & Haro 4-145 & 2.57$\pm$0.08 & 1.34$\pm$0.07 & $-$1.44$\pm$0.06 & 2.77\\
77 & 3017260915384911616 &  II Ori & 2.56$\pm$0.02 & 1.31$\pm$0.02 & $-$1.65$\pm$0.01 & 1.32 \\
78 & 3017161753180173184 &  V937 Ori & 2.56$\pm$0.06 & 1.43$\pm$0.06 & 0.98$\pm$0.05 & 1.04 \\
79 & 3017138663436006400 &  V2029 Ori & 2.55$\pm$0.03 & 1.15$\pm$0.03 & $-$2.10$\pm$0.02 & 1.06 \\
80 & 3016548431848965888 &  NT Ori & 2.56$\pm$0.02 & 0.41$\pm$0.02 & 0.01$\pm$0.02 & 1.21\\
81 & 3016952090054853888 &  V571 Ori & 2.70$\pm$0.03 & 0.65$\pm$0.03 & 1.46$\pm$0.03 & 2.35 \\
82 & 3015798633638359936 & Haro 4-488 & 2.34$\pm$0.03 & 0.04$\pm$0.02 & $-$1.06$\pm$0.02 & 1.05 \\
83 & 3209529180840270720 &  V415 Ori & 2.51$\pm$0.01 & 0.34$\pm$0.01 & 0.05$\pm$0.01 & 1.05 \\
84 & 3012179178798021248 & 2MASS J05404851-0923468 & 2.38$\pm$0.08 & $-$0.51(0.07 & $-$1.78$\pm$0.07 & 3.06\\
85 & 3016953601883341312 &  V814 Ori & 2.58$\pm$0.04 & 0.67$\pm$0.03 & 1.32$\pm$0.03 & 1.14 \\
86 & 3017399522579317248 &  V657 Ori & 2.52$\pm$0.08 & 2.32$\pm$0.08 & $-$0.35$\pm$0.06 & 1.99 \\
87 & 3016935563020738688 &  V2720 Ori & 2.55$\pm$0.03 & 0.43$\pm$0.03 & 0.21$\pm$0.03 & 0.91 \\
88 & 3017266481662489984 &  V1481 Ori & 2.54$\pm$0.03 & 1.04$\pm$0.02 & 0.23$\pm$0.02 & 1.27 \\
89 & 3016535680090105344 & UCAC4 415-010131 & 2.62$\pm$0.04 & 1.12$\pm$0.04 & 0.24$\pm$0.03 & 1.07 \\
90 & 3015649443653596288 & Haro 4-446 & 2.44$\pm$0.05 & $-$0.41$\pm$0.04 & $-$0.17$\pm$0.04 & 2.10 \\
91 & 3209584671817511040 & 2MASS J05353976-0440242 & 2.40$\pm$0.04 & 1.50$\pm$0.03 & $-$0.55$\pm$0.02 & 1.42 

\label{tab_app:gaia_info}
\end{longtable}
\tablefoot{$^{*}$ Sources with RUWE $>$ 2.5 \citep{Fitton_2022} and {\tt fidelity\_v2} $<$ 0.5 \citep{Rybizki_2022}.}

\newpage

\section{Stellar and accretion properties derived}\label{app:add_tables}

The stellar and accretion properties derived with \texttt{FRAPPE} are presented in Table~\ref{tab:fitter_output}.

\renewcommand{\arraystretch}{1.15}

\begin{table*}[h!]
\caption{Literature parameters and parameters derived with \texttt{FRAPPE} for the VLT/X-Shooter sample.}
\begin{center}
\resizebox{\textwidth}{!}{
\begin{tabular}{llllccccccccccc}
\hline \hline
\multirow{2}{*}{ID} & \multirow{2}{*}{OmegaCAM} & Lit. & \multirow{2}{*}{Ref.$^{a}$} & \multirow{2}{*}{SpT} & \Teff  & $A_V$ & \multirow{2}{*}{log(\Lstar/\Lsun)} & \multirow{2}{*}{log(\Lacc/\Lsun)} & \multirow{2}{*}{\Rstar [$R_{\odot}$]} & \multicolumn{2}{c}{PISA} & \multicolumn{2}{c}{Baraffe+2015} & Accretion \\ 
 &  & SpT &  &  & [K] & [mag] &  &  &  & \Mstar [$M_{\odot}$] & log\Macc [\Msun\,yr$^{-1}$] & \Mstar [$M_{\odot}$] & log\Macc [\Msun\,yr$^{-1}$] & upper limit? \\ \hline
1 & 1042831 &  &  & K7.5 & 3960 & 1.4 & $-$1.19 & $-$0.02        & 2.08$_{-0.41}^{+0.62}$ & 0.54$_{-0.07}^{+0.09}$ & $-$8.00$_{-0.30}^{+0.27}$ & 0.61$_{-0.08}^{+0.09}$ & $-$8.06$_{-0.30}^{+0.27}$ &  \\
2 & 1021191 & M2.5 & H97 & M1.5 & 3640 & 1.0 & $-$1.91 & 0.0   & 2.50$_{-0.47}^{+0.74}$ & 0.34$_{-0.03}^{+0.03}$ & $-$8.45$_{-0.29}^{+0.26}$ & 0.39$_{-0.03}^{+0.03}$ & $-$8.51$_{-0.29}^{+0.26}$ & Y \\
3 & 1079659 & M0.0 & K16 & M1.0 & 3720 & 0.3 & $-$1.15 & $-$0.62 & 1.18$_{-0.22}^{+0.35}$ & 0.49$_{-0.05}^{+0.06}$ & $-$8.17$_{-0.29}^{+0.27}$ & 0.51$_{-0.05}^{+0.06}$ & $-$8.18$_{-0.29}^{+0.26}$ &  \\
4 & 1035737 & K7.0 & K16 & K6.5 & 4067 & 1.6 & $-$1.17 & $-$0.34 & 1.37$_{-0.26}^{+0.41}$ & 0.74$_{-0.08}^{+0.07}$ & $-$8.31$_{-0.29}^{+0.26}$ & 0.76$_{-0.07}^{+0.06}$ & $-$8.32$_{-0.29}^{+0.26}$ &  \\
5 & 996321 & K7.5 & K16 & K6.5 & 4067 & 2.7 & $-$1.17 & 0.08   & 2.22$_{-0.42}^{+0.66}$ & 0.60$_{-0.06}^{+0.09}$ & $-$8.00$_{-0.29}^{+0.27}$ & 0.69$_{-0.06}^{+0.07}$ & $-$8.06$_{-0.29}^{+0.26}$ &  \\
6 & 1031924 & M2.5 & DR10 & M1.5 & 3640 & 0.8 & $-$1.77 & $-$0.58& 1.28$_{-0.24}^{+0.38}$ & 0.42$_{-0.04}^{+0.05}$ & $-$8.69$_{-0.29}^{+0.27}$ & 0.44$_{-0.04}^{+0.05}$ & $-$8.70$_{-0.29}^{+0.27}$ &  \\
7 & 1056861 & M2.0 & K16 & M3.0 & 3410 & 0.2 & $-$1.71 & $-$0.57 & 1.49$_{-0.29}^{+0.44}$ & 0.27$_{-0.03}^{+0.04}$ & $-$8.37$_{-0.29}^{+0.27}$ & 0.29$_{-0.03}^{+0.03}$ & $-$8.40$_{-0.29}^{+0.26}$ &  \\
8 & 1051021 & M2.5 & H97 & M1.5 & 3640 & 2.1 & $-$1.27 & $-$0.23 & 1.93$_{-0.37}^{+0.57}$ & 0.37$_{-0.03}^{+0.04}$ & $-$7.95$_{-0.29}^{+0.26}$ & 0.40$_{-0.03}^{+0.03}$ & $-$7.98$_{-0.29}^{+0.26}$ &  \\
10 & 985345 & M2.5 & H97 & M1.5 & 3640 & 0.1 & $-$1.49 & $-$0.75 & 1.07$_{-0.20}^{+0.32}$ & 0.44$_{-0.04}^{+0.06}$ & $-$8.50$_{-0.29}^{+0.27}$ & 0.46$_{-0.05}^{+0.05}$ & $-$8.52$_{-0.29}^{+0.26}$ &  \\
... & ... & ... & ... & ... & ... & ... & ... & ... & ... & ... & ... & ... & ... &  \\
\hline
\end{tabular}}
\tablefoot{$^{a}$ CK79: \citet{Cohen_Kuhi_1979}; H97: \citet{Hillenbrand_1997}; P09: \citet{Parihar_2009}; F09: \citet{Fang_2009}; DR10: \citet{DaRio_2010}; CoG12: \citet{CarattioGaratti_2012}; H12: \citet{Hsu_2012}; F13: \citet{Fang_2013}; K16: \citet{Kim_2016}; K17: \citet{Kounkel_2017}; F17: \citet{Fang_2017}; B19: \citet{Briceno_2019}; B20: \citet{Birky_2020}; Z23: \citet{Zhang_2023}.}
\end{center}
\label{tab:fitter_output}
\end{table*}

\vspace{-25pt}
\section{Implementation of \texttt{STAR-MELT} in sub-luminous sources}

\begin{figure}[h!]
    \centering

    \makebox[\textwidth][c]{%
        \includegraphics[width=0.35\textwidth]{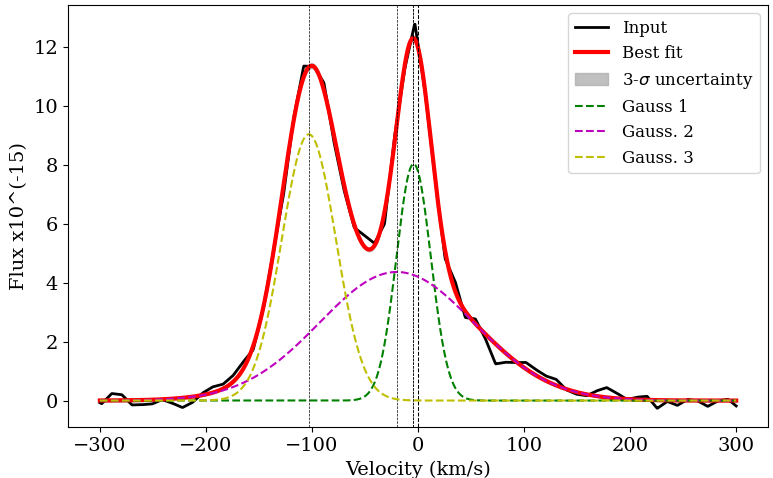}\hspace{0.05\textwidth}
        \includegraphics[width=0.35\textwidth]{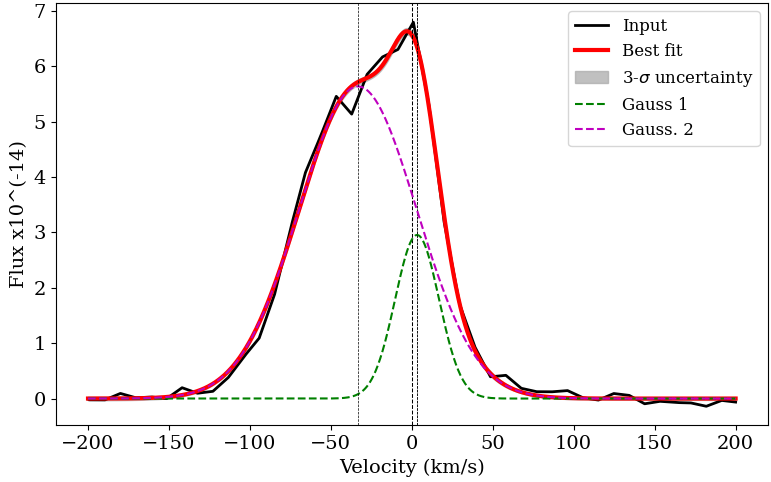}
    }

    \vspace{0.1cm}

    \makebox[\textwidth][c]{
        \includegraphics[width=0.35\textwidth]{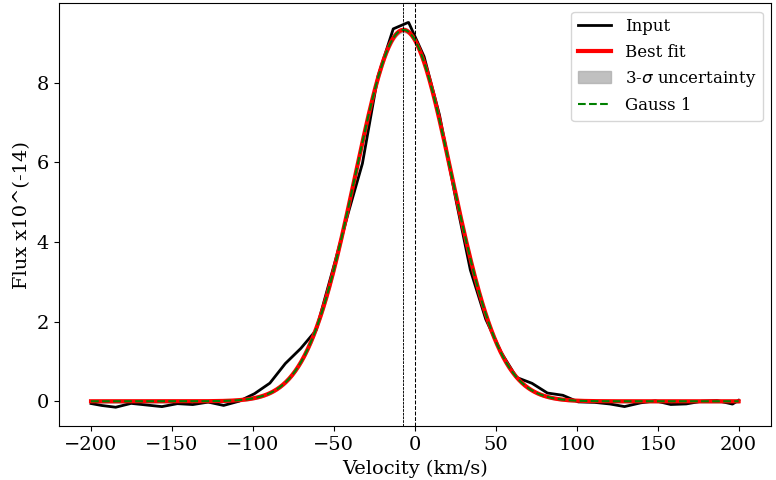}\hspace{0.05\textwidth}
        \includegraphics[width=0.35\textwidth]{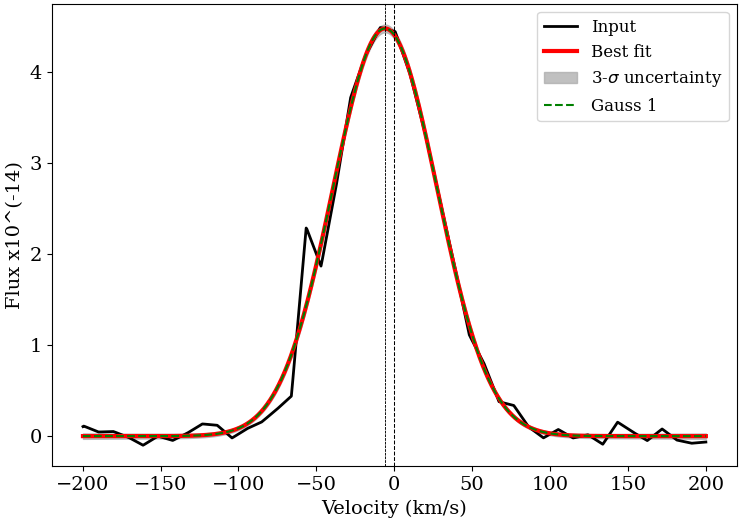}
    }

    \caption{Best fit from \texttt{STAR-MELT} \citep{Campbell-White_2021} for the [\ion{O}{i}]\,$\lambda$6300 \AA\ line for the sub-luminous sources discussed in Sect.~\ref{subsec:sub-luminous}. From top left to bottom right the ID sequence is the following \#38, \#39, \#48, and \#57. Sources in the top row are fitted with more than one Gaussian, whereas those in the bottom row are better reproduced with a single Gaussian. The best fit is shown as a solid red line. All lines are corrected for the stellar radial velocity.}
    \label{fig:star-melt_subl}
\end{figure}

As described in Sect.~\ref{subsec:sub-luminous}, we implemented the multi-Gaussian fitting procedure from the \texttt{STAR-MELT} package \citep{Campbell-White_2021} on the [\ion{O}{i}] line at $\lambda$630\,nm of the identified sub-luminous sources. We show the resulting best fits in Fig. \ref{fig:star-melt_subl}. Only the low-velocity components (|v|$<$40\,km\,s$^{-1}$) were considered in the conversion to \Lacc. For that, we used the empirical relation presented by \citet{Nisini_2018}.

\vspace{-25pt}
\section{Comparison of accretion properties between the extended ONC and L1641/L1647}\label{app:regions_scaling_relations}

\begin{figure*}[h!]
    \centering
    \includegraphics[width=0.45\textwidth]{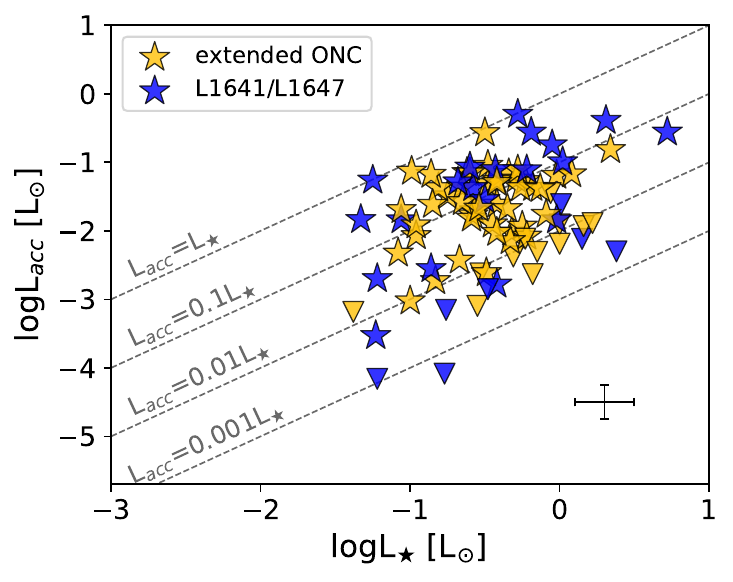}
    \includegraphics[width=0.45\textwidth]{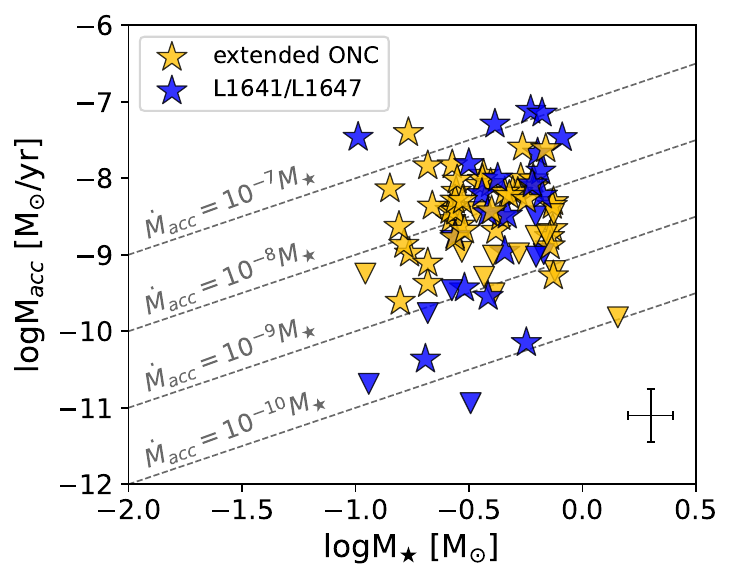}
    \caption{$L_\mathrm{acc}$\,$-$\,$L_\star$ (left) and $\dot{M}_\mathrm{acc}$\,$-$\,$M_\star$ (right) relations for sources in the extended ONC (yellow stars) and L1641/L1647 (blue stars). Downward triangles indicate accretion upper limits (Sect.~\ref{subsec:acc_limits}). Both populations occupy the same parameter space with no evident systematic differences.}
    \label{fig:regions}
\end{figure*}

In Fig.~\ref{fig:regions}, we show the \Lacc–\Lstar\ and \Macc–\Mstar\ relations separated by sub-region within Orion A, distinguishing between the extended ONC and L1641/L1647. The two populations largely overlap in both diagrams, suggesting that accretion properties do not strongly depend on the sub-region. L1641/L1647 sources display a somewhat larger spread, particularly in the \Macc–\Mstar\ diagram, where a handful of sources exhibit low accretion rates. Given that both populations are confined to a narrow stellar mass range (0.1$-$1 M$_{\odot}$), we do not attempt to constrain the power-law slope of the \Macc–\Mstar\ relation for each sub-region independently.

\section{Disk dust masses}\label{app:dust_masses}

We retrieved available dust masses for our sample from several works: 1) \citet{vanTerwisga_2019} measured dust masses of disks in the Orion Molecular Cloud 2 (OMC-2) region from 3\,mm ALMA continuum observations. By considering a matching radius of 0.5'', we found that 2 sources in our sample have dust mass estimates from this work; 2) \citet{vanTerwisga_2022} conducted the SODA survey, the largest single ALMA disk dust mass survey in the southern regions of Orion A (L1641 and L1647 clouds) in Band 6. Since we designed our observational campaign to have a full overlap with SODA, all of the sources with declinations $<-$6$\degree$ (i.e. in the L1641/L1647 regions) have an ALMA counterpart from this work, resulting in 27 additional dust masses; then, 3) \citet{Flaherty_2025} surveyed over 100 spectroscopic tight binaries in Orion A with ALMA Band 6 and we found that three sources from our sample are included in their catalog. There exist additional ALMA continuum observations in the Orion A complex \citep[e.g.][]{Mann_2014,Eisner_2018,Grant_2021,Otter_2021,Ballering_2023}, but these surveys target compact regions of the complex, typically the Trapezium cluster, which are not covered by our VLT/X-Shooter sample.

\begin{figure}[h!]
    \centering
    \includegraphics[width=0.58\textwidth]{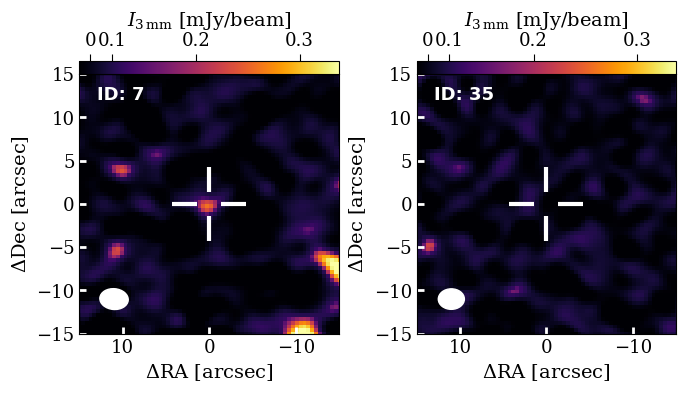}
    \caption{Cutouts of the 3\,mm continuum ALMA maps from the EMERGE project~\citep{Hacar_2024} around the positions of sources \#7 (left) and \#35 (right) in the VLT/X-Shooter survey. White circles indicate the ALMA beam; the position of the target is marked with a white crosshair symbol.}
    \label{fig:ALMAstamps}
\end{figure}

Two more sources (IDs \#7 and \#35, [MGM2012] 2300 and [MGM2012] 1162 respectively) were covered by 3\,mm continuum maps of the OMC-3 and OMC-4 fields that are part of the EMERGE program, and were published in~\citet{Hacar_2024}. Of these two sources, source \#7 ([MGM2012] 2300) is detected at $4.7\sigma$. These ALMA observations are part of the same project used to extract continuum fluxes of disks in OMC-2 presented by~\citet{vanTerwisga_2019}. They are low-resolution ($\sim3''$ beam) observations covering a large area. To measure the disk flux, we extracted the peak flux in a $3''$ radius around the source position, estimating the background noise in a $6''$ radius annulus around it. To make sure the fluxes are not contaminated by large-scale emission from the nearby Integral Shaped Filament (ISF), we performed an additional visual inspection to ensure that no extended emission was contaminating the image, using Fig.~\ref{fig:ALMAstamps}. The 3\,mm fluxes were converted to dust masses, assuming no significant free-free contamination (given the relatively low UV background in these fields) and following the same opacities and disk temperature assumptions as in~\citet{vanTerwisga_2019}. With this, we compiled dust mass measurements for 34 out of the total 91 targets. Table \ref{tab:Mdust_lit} summarizes the available measurements.

\renewcommand{\arraystretch}{1.2}
\begin{table}[h!]
\caption{Dust disk mass estimates available in the literature for our sample.}
\begin{center}
\scalebox{0.85}{
\begin{tabular}{lcccc}
\hline \hline
ID & F$_{\rm mm}$\,[mJy] & $M_{dust}$\,[M$_{\oplus}$] & Band & Ref. \\
\hline
7 & 0.25$\pm$0.05 & 14.6$\pm$3.1 & 3 & 4 \\
20 & $<$0.46$\pm$0.27 & $<$12.30$\pm$15.51 & 3 & 1 \\
27 & 0.95$\pm$0.12 & 40.50$\pm$6.64 & 3 & 1 \\
34 & 3.19$\pm$0.07 & 5.79$\pm$0.18 & 6 & 3 \\
35 & $<$0.12$\pm$0.07 & $<$7.0$\pm$4.1 & 3 & 4 \\
38 & 0.64$\pm$0.09 & 2.90$\pm$0.39 & 6 & 2 \\
39 & 6.07$\pm$0.10 & 27.40$\pm$0.43 & 6 & 2 \\
41 & 10.57$\pm$0.08 & 51.90$\pm$0.41 & 6 & 2 \\
43 & 75.14$\pm$0.11 & 363.90$\pm$0.55 & 6 & 2 \\
46 & 0.86$\pm$0.08 & 3.90$\pm$0.35 & 6 & 2 \\
51 & 0.50$\pm$0.09 & 2.20$\pm$0.40 & 6 & 2 \\
52 & 0.73$\pm$0.08 & 3.30$\pm$0.34 & 6 & 2 \\
53 & 18.23$\pm$0.10 & 82.10$\pm$0.44 & 6 & 2 \\
55 & 5.13$\pm$0.06 & 28.90$\pm$0.31 & 6 & 2 \\
56 & 1.64$\pm$0.06 & 8.40$\pm$0.31 & 6 & 2 \\
58 & 22.96$\pm$0.09 & 103.50$\pm$0.41 & 6 & 2 \\
62 & 3.31$\pm$0.08 & 11.10$\pm$0.83 & 6 & 3 \\
63 & $<$0.34$\pm$0.09 & $<$1.50$\pm$0.39 & 6 & 2 \\
64 & 8.21$\pm$0.08 & 37.00$\pm$0.35 & 6 & 2 \\
65 & 2.37$\pm$0.10 & 14.20$\pm$0.57 & 6 & 2 \\
66 & 4.55$\pm$0.06 & 21.50$\pm$0.26 & 6 & 2 \\
67 & 3.25$\pm$0.07 & 14.60$\pm$0.31 & 6 & 2 \\
75 & 1.11$\pm$0.09& 5.00$\pm$0.38 & 6 & 2 \\
78 & 2.09$\pm$0.09 & 9.50$\pm$0.38 & 6 & 2 \\
79 & 1.67$\pm$0.07 & 7.60$\pm$0.33 & 6 & 2 \\
80 & 12.29$\pm$0.08 & 55.50$\pm$0.36 & 6 & 2 \\
81 & 3.11$\pm$0.08 & 14.00$\pm$0.35 & 6 & 2 \\
82 & 4.40$\pm$0.07 & 22.00$\pm$0.33 & 6 & 2 \\
84 & 0.52$\pm$0.08 & 3.1$\pm$0.50 & 6 & 2 \\
85 & 1.63$\pm$0.08 & 7.3$\pm$0.36 & 6 & 2 \\
87 & 3.27$\pm$0.08 & 14.8$\pm$0.36 & 6 & 2 \\
88 & 1.73$\pm$0.07 & 5.75$\pm$0.26 & 6 & 3 \\
89 & 0.38$\pm$0.07 & 1.7$\pm$0.33 & 6 & 2 \\
90 & 1.07$\pm$0.06 & 5.2$\pm$0.31 & 6 & 2 \\
\hline
\end{tabular}
}
\tablefoot{\textbf{References:} 1: \citet{vanTerwisga_2019}; 2: \citet{vanTerwisga_2022}; 3: \citet{Flaherty_2025}; 4 $-$ This work.}
\end{center}
\vspace{-15pt}
\label{tab:Mdust_lit}
\end{table}

\end{appendix}

\end{document}